\documentclass[11pt]{article}

\usepackage{amsfonts,amssymb,amsbsy,latexsym,amsmath,tabulary,graphicx,times,xcolor}
\usepackage[top=50mm, bottom=50mm, left=50mm, right=50mm]{geometry}
\usepackage{lineno}
\usepackage{amssymb}
\usepackage{amsmath}
\usepackage{amsthm}
\usepackage{epsfig}
\usepackage{graphicx}
\usepackage{float}
\usepackage{multirow}
\usepackage{color}
\usepackage{lineno}
\usepackage{fullpage}
\usepackage[normalem]{ulem} 
\usepackage{makeidx}
\usepackage{xspace}
\usepackage{wrapfig}
\makeindex

\usepackage{flafter}

\usepackage{dblfloatfix}
\usepackage{blindtext}

\definecolor{darkred}{rgb}{1, 0.1, 0.3}
\definecolor{darkblue}{rgb}{0.1, 0.1, 1}
\definecolor{darkgreen}{rgb}{0,0.6,0.5}

\newcommand {\mm}[1] {\ifmmode{#1}\else{\mbox{\(#1\)}}\fi}

\usepackage{esvect}
\usepackage{tabulary,xcolor}
\usepackage{amsfonts,amsmath,amssymb}
\usepackage{siunitx}

\usepackage{ifluatex}
\ifluatex
\usepackage{fontspec}
\defaultfontfeatures{Ligatures=TeX}
\usepackage[]{unicode-math}
\unimathsetup{math-style=TeX}
\else 
\usepackage[utf8]{inputenc}
\fi 
\ifluatex\else\usepackage{stmaryrd}\fi

\usepackage{url,multirow,morefloats,floatflt,cancel,textcomp,tfrupee}
\usepackage{pifont}
\usepackage[nointegrals]{wasysym}
\usepackage{tikz}
\usepackage{graphicx}
\usetikzlibrary{positioning}

\usepackage{url,hyperref,lineno,microtype,subcaption}
\usepackage[onehalfspacing]{setspace}

\usepackage[title]{appendix}
\usepackage{algorithm}
\usepackage[noend]{algpseudocode}
\usepackage{algorithmicx}

\algdef{SE}[DOWHILE]{Do}{doWhile}{\algorithmicdo}[1]{\algorithmicwhile\ #1}%

\makeatletter
\def\BState{\State\hskip-\ALG@thistlm}
\makeatother

\usepackage{authblk}

\begin{document}

\title{Robust Monte-Carlo Simulations in Diffusion-MRI: Effect of the substrate complexity and parameter choice on the reproducibility of results}

\author[1]{Jonathan Rafael-Patino}
\author[1]{David Romascano}
\author[2]{Alonso Ramirez-Manzanares}
\author[3,4,5]{Erick Jorge Canales-Rodríguez}
\author[3,1]{Gabriel Girard}
\author[1,3,5]{Jean-Philippe Thiran}
\affil[1]{Signal Processing Lab (LTS5)\unskip, \'{E}cole Polytechnique F\'{e}d\'{e}rale de Lausanne\unskip, Lausanne\unskip, Switzerland}
\affil[2]{Centro de Investigaci\'{o}n en Matem\'{a}ticas\unskip, Guanajuato\unskip, Gto\unskip, M\'{e}xico}
\affil[3]{Radiology Department\unskip, Centre Hospitalier Universitaire Vaudois\unskip, Lausanne\unskip, Switzerland}
\affil[4]{Centre Hospitalier Universitaire Vaudois\unskip, Lausanne\unskip, Switzerland}
\affil[5]{FIDMAG Germanes Hospitalàries, Sant Boi de Llobregat\unskip, Barcelona\unskip, Spain}
\affil[6]{Mental Health Research Networking Center (CIBERSAM)\unskip, Madrid\unskip, Spain}
\affil[7]{University of Lausanne \unskip, Lausanne\unskip, Switzerland.}




\maketitle
\setcounter{page}{0}

\def\checkGraphicsWidth{\ifdim\Gin@nat@width>\textwidth
	\tsGraphicsScaleX\textwidth\else\Gin@nat@width\fi}

\def\checkGraphicsHeight{\ifdim\Gin@nat@height>.9\textheight
	\tsGraphicsScaleY\textheight\else\Gin@nat@height\fi}

\def\fixFloatSize#1{\@ifundefined{processdelayedfloats}{\setbox0=\hbox{\includegraphics{#1}}\ifnum\wd0<\columnwidth\relax\renewenvironment{figure*}{\begin{figure}}{\end{figure}}\fi}{}}

\begin{abstract}

Monte-Carlo Diffusion Simulations (MCDS) have been used extensively as a ground truth tool for the validation of microstructure models for Diffusion-Weighted MRI. However, methodological pitfalls in the design of the biomimicking geometrical configurations and the simulation parameters can lead to approximation biases. Such pitfalls affect the reliability of the estimated signal, as well as its validity and reproducibility as ground truth data. In this work, we first present a set of experiments in order to study three critical pitfalls encountered in the design of MCDS in the literature, namely, the number of simulated particles and time steps, simplifications in the intra-axonal substrate representation, and the impact of the substrate's size on the signal stemming from the extra-axonal space. The results obtained show important changes in the simulated signals and the recovered microstructure features when changes in those parameters are introduced. Thereupon, driven by our findings from the first studies, we outline a general framework able to generate complex substrates. We show the framework’s capability to overcome the aforementioned simplifications by generating a complex crossing substrate, which preserves the volume in the crossing area and achieves a high packing density. The results presented in this work, along with the simulator developed, pave the way towards more realistic and reproducible Monte-Carlo simulations for Diffusion-Weighted MRI.

\end{abstract}

\newpage

\section{Introduction}

Diffusion-Weighted Magnetic Resonance Imaging (DW-MRI) is a non-invasive technique with enormous potential for the study of the brain's microstructure by  measuring the diffusion properties of biological tissue. For instance, state-of-the-art methods can use these measurements to estimate tissue properties of the brain white matter, e.g. axonal diameter estimations\unskip~\cite{72980:2907595}, orientation and volume fraction of the axonal bundles\unskip~\cite{72980:2907716,Zhang2011} and neurite dispersion\unskip~\cite{72980:2907515}. The previous information is valuable to understand the brain's maturation\unskip~\cite{72980:2907473,Sexton2014AcceleratedCI} as well as the degeneration process associated with neuronal diseases like axonal degeneration\unskip~\cite{72980:2907872} and multiple sclerosis\unskip~\cite{Trapp1998}.

In DW-MRI, an attenuated signal is usually recovered via a Pulsed Gradient Spin-Echo protocol (PGSE)~\cite{PGSE} sensitive to the displacement of the water molecules. Analytical solutions of the signal attenuation can be derived for simple geometrical shapes such as impermeable planes, cylinders, and spheres~\unskip~\cite{72980:2907502}. However, for applications where the signal attenuation of complex cellular structures or non-homogeneous media is needed, e.g. to generate ground truth data, an analytical solution is no longer feasible to pursue due to its inherent complexity. Because of this, simplifications of the diffusion media have been used as the backbone of most of the microstructure models in the literature\unskip~\cite{72980:2907503,72980:2907601,72980:2907515,Ferizi2015}.

Monte-Carlo Diffusion Simulations (MCDS) provide a fundamental approach to study the diffusion phenomena in scenarios where the analytical solutions cannot be computed due to their complexity. In contrast with other numerical methods, MCDS does not require an explicit model of the diffusion signal for a given geometry. Instead, MCDS require an accurate geometrical and physical representation of the diffusion media (called the substrate), a large number of samples, and an acquisition protocol like the classical Pulsed Gradient Spin-Echo (PGSE). In general, an accurate approximation of the diffusion signals, mimicking the ones obtained from the brain's white matter, can be computed if the substrate captures the relevant white matter microstructure features and the simulation parameters are tuned properly (number of samples and their step sizes). Despite this, many simplifications are usually used in order to decrease the computational burden. The most common ones include the use of substrates of small size, the use of a limited number of samples, the use of simple geometries, and the restriction of the 3D diffusion to the 2D case~\unskip~\cite{72980:2907552,72980:2907499,72980:2907613,72980:2907483}.


\cite{72980:2907499} presented the first work, to the best of our knowledge, which employs MCDS to study the extracellular diffusion in brain tissue. In this work, 2D histological data was used to draw binary contours to be used as irregular intracellular barriers. From this study on, most studies simplified the representation of the extracellular space as a collection of restricted corridors in the orthogonal plane of the axonal direction and unrestricted parallel to them\unskip~\cite{Deriche1,72980:2907614,72980:2907613,72980:2907607}. In addition, the intracellular compartment is usually idealised as a collection of parallel hollow cylinders with constant radii or radii sampled from a distribution estimated from histological data\unskip~\unskip~\cite{RAFFELT, 72980:2907563, 72980:2907481, 72980:2907508,72980:2907481}. Recent studies have suggested that such simplification cannot capture the complexity of the axonal structures of white matter, and thus its diffusion characteristics\unskip~\cite{72980:2907485,Ginsburger2018}. For instance, changes in the diffusion signal and parameters derived from the diffusion tensor, such as the fractional anisotropy (FA) and mean diffusivity (MD), were obtained by introducing regular undulations in the intra-axonal compartment \unskip~\cite{72980:2907473}.    

Because of the aforementioned problems, a number of works proposed experiments where non-trivial structures were used as intracellular substrates, e.g. dispersed axons\unskip~\cite{Ginsburger2018}, the presence of abutting cylinders \unskip~\cite{72980:2907482} and arbitrarily generated meshes \unskip~\cite{72980:2907480}. However, to this day, such approaches have not been thoroughly adopted by the DW-MRI community because of the high computational burden they demand and the lack of available tools. Because of this, more realistic diffusion simulations remain virtually unexploited. 

In this work, we study three important pitfalls encountered in the design of MCDS in the literature used to reduce the computational burden, namely the number of simulated particles and the number of time steps, the intracellular geometrical representation, and the generated extra-axonal space in terms of the substrate's size. Each experiment presented below illustrates a possible bias induced in the computed signal when such simplifications are not properly addressed, which affects its reproducibility. Finally, driven by the results from the previous experiments, we outline a general framework that can be used to generate complex substrates in order to overcome the limitations of previous studies.

\section{Theory}

The obtained signal from a PGSE DW-MRI measurement, at a time $t$,  is given by\unskip~\cite{72980:2907471}

\begin{equation}
S(t=TE) = S_0 \int P(\phi,t) e^{-i\phi}d\phi ,
\label{eq:obtainedSignal}
\end{equation}
where $S_{0}$ denotes the signal obtained in the absence of a diffusion gradient magnetic field, $TE$ is the echo time,  $P(\phi,t)$ is the phase distribution function  of the spin ensemble at time $t = TE$, and $\phi$ is the accumulated phase shift of the spin.

The amount of attenuation of a single diffusing spin on the measured PGSE signal is proportional to the dephasing due to the effect of the time-dependent magnetic field $G(t)$, and the spin's displacement. For a single spin and a given magnetic gradient vector, the phase shift due to the applied gradient over time can be numerically formulated as in \unskip~\cite{72980:2907471}

\begin{equation}
\phi(t)=a(t)\gamma G(t)\cdot x(t) ,
\label{eq:local_phase_shift}
\end{equation}
where $\gamma$ is the gyromagnetic ratio, $G(t)$ is the applied magnetic diffusion gradient at time $t$, $x(t)$ is the spin's displacement from the starting position, and $a(t)$ is a function that shifts the sign of the gradient vector due to the refocusing Radio Frequency (RF) pulse. In a classic PGSE experiment: $a(t)$ is equal to $+1$ for all time $t$ before the RF pulse and $-1$ after. The produced attenuated signal is then the result of the accumulated phase shift of the full assembly of spins at the TE, given by

\begin{equation}
S(t=TE)/S_0=\left\langle  e^{{-i\int^{t=TE}_{0}}\phi(t)dt'}\right\rangle .
\label{eq:ensembled_signal}
\end{equation}

\subsection{Simulation fundamentals}

Equation~\eqref{eq:ensembled_signal} can be approximated using a finite number of spin samples $N_s$ over a discrete time lapse, following an approach as in\unskip~\cite{72980:2907483}:

\begin{equation} 
S/S_0=\frac1{N_s}\sum^{N_s}e^{-i\sum^{Nt}_{t}\phi(t) dt} ,
\label{eq:numercial_decay}
\end{equation}
where $dt$ is the step duration, defined as the total diffusion time divided by the number of steps taken (N\ensuremath{_{t}}). The value of $dt$ can be fixed as in\unskip~\cite{72980:2907508,72980:2907482} or normally distributed as in\unskip~\cite{72980:2907474}. In our exploration, we made use of fixed step sizes derived from Einstein's equation: $r=\sqrt{(6Ddt)}$, where $D$ is the diffusion coefficient and $r$ is the expected mean displacement. A fixed step size have been shown by~\unskip~\cite{72980:2907508} to "reduce the fluctuation in the mean-square displacement of the spins and improve the convergence in the model". Moreover,~\unskip~\cite{72980:2907478} have shown the fixed step size to be better suited for non-homogeneous systems.

The idea behind a Monte-Carlo Diffusion Simulator is to compute equation~\eqref{eq:numercial_decay} by simulating the particles' Brownian motion and their interaction with respect to a defined substrate. At the beginning of the simulation, the particles are uniformly placed inside the defined substrate's voxel, or substrate's limits. This way, the number of particles in all compartments is proportional to the defined volume fractions. If necessary, the local position of each particle can be tracked to separate the signal contribution of each compartment by, for example, tracking if the particle is inside a given compartment. Over the duration of the simulation, the simulated particles collide and bounce with the substrate's barriers, depending on the barrier properties. Finally, the accumulated phase shift is tracked depending on the spin-echo protocol using equation~\eqref{eq:numercial_decay}. 

Overall, the formulation above presents an accurate numerical approximation of the diffusion signal based solely in the phase shift distribution. However, is worth noticing that many other effects such as noise levels or the magnetization relaxation should be considered in order to approximate a more realistic DWI-MRI signal.

\section{Material and Methods}

All the simulated signals presented below were computed using the sum of the accumulated phase shift approximation showed in Equation~\ref{eq:numercial_decay} implemented in an in-house Monte-Carlo simulator. The simulator employs a similar approach to compute the diffusion signal as the ones presented in \unskip~\cite{72980:2907508,72980:2907482}. The simulator uses a hybrid GPU/Multi-CPU framework, implemented in C++11. It includes routines to optimise the collision detection and the memory consumption based on the complexity analysis of Appendix~\ref{App:B}; making the software able to handle simulations of 3D meshed substrates with millions of triangles and particles. The simulator was initially validated by verifying that the generated signals from particles within impermeable planes, cylinders, and spheres were equal to those obtained from their corresponding analytical solutions. Moreover, results in more complex domains including the extra-axonal space of brain tissue, were comparable to those obtained from an alternative and independent Finite Element Method approach described in~\cite{proc-ISMRM-FEM}. The substrates' data, meshes, and the simulator are available from the corresponding author upon request on the paper's Git-Hub repository: (\url{https://github.com/jonhrafe/Robust-Monte-Carlo-Simulations}).

\subsection{Confidence level estimation}
\label{Sec:Conf}

In Monte-Carlo based methods, the number of samples is critical for the confidence level of the estimated results. However, the number of particles has the most significant impact on the computational burden. To highlight the importance of the number of simulated spins, an experiment was performed in order to quantify the variance of the estimated signal as a function of the number of particles sampled in a substrate. To do this, the errors of a set of simulated signals with different numbers of samples were measured. The measured errors were compared against the expected analytical solution in the intra-axonal space and for a gold-standard estimation of the extracellular space. A substrate with $10,000$ parallel cylinders with diameters sampled from a Gamma distribution, $\Gamma (\kappa, \theta$), with shape, $\kappa = 4.0$, and scale, $\theta=4.5\times 10^{-7}$, was used, resulting in a mean diameter $\mu$ = \SI{1.8}{\micro\meter} with a standard deviation of $\sigma$ = \SI{0.9}{\micro\meter}, using a packing algorithm as the one described in~\cite{72980:2907508}, which results in a distribution of radii comparable to the ones found in the literature~\cite{Zhang2011, Benjamini2016, 72980:2907613}.

This substrate was used since the analytical signal of the intra-axonal space can be computed using the volume-weighted sum of the individual signals:

\begin{equation}
S_i = \frac{ v_1 Sc_{i,1} + \dots + v_n Sc_{i,n}}{\sum v_j}, 
\end{equation}
where $S_i$ is the $i^{th}$ acquisition, $v_j$ is the volume of the $j^{th}$ cylinder and $Sc_{i,1}$ is the analytical signal of the cylinder obtained using the Gaussian Phase Distribution (GPD) approximation of the signal in cylinders for a given radius~\cite{VanGelderen1994}. Figure \ref{fig:boots_gamma_signal} shows the resulting distribution of radii as well as the computed ground-truth intra-axonal signal. For the extracellular signal, as there is no analytical model, the gold-standard was estimated using a very high number of particles: $20\times 10^6$ particles, and time-steps: $2 \times 10^4 $ steps. These parameters were chosen based on previous results~\cite{proc-ISMRM-FEM} and by studying the convergence properties for higher numbers of particles and time-steps~\cite{72980:2907508}. In fact, we verified that the signal converges for even less demanding simulation parameters (i.e., $1\times 10^6$ particles, and $5\times 10^3$ steps). In order to keep results as accurate as possible, however, we decided to use simulation parameters higher than the minimum required.

The estimated signals were computed varying the number of particles from $1 \times 10^3$ to $1\times 10^6$ particles, and the time-steps from $1\times 10^2$ to $2\times 10^4$ steps. The diffusion coefficient was fixed to $D = 0.6  \times 10^{-3}$ \SI{}{mm^2/s} (corresponding to an ex-vivo diffusivity), and TE = $0.054$~\SI{}{s}, for both, the simulations and the ground-truth data. The original optimized ActiveAx PGSE protocol~\cite{72980:2907481} was used, which consist of a four shell HARDI acquisition with $90$ orientations per shell, each shell with the following parameters respectively, i) $b=1930$~\SI{}{s/mm^2}, $G = 140$~\SI{}{mT/m}, $\delta = 0.010$~\SI{}{s}, and $\Delta = 0.016$~\SI{}{s}; ii) $b=1930$~\SI{}{s/mm^2}, $G = 140$~\SI{}{mT/m}, $\delta = 0.010$~\SI{}{s}, and $\Delta = 0.016$~\SI{}{s}; iii) $b=3090$~\SI{}{s/mm^2}, $G = 131$~\SI{}{mT/m}, $\delta = 0.007$~\SI{}{s}, and $\Delta = 0.045$~\SI{}{s}; iv) $b=13190$~\SI{}{s/mm^2}, $G = 140$~\SI{}{mT/m}, $\delta = 0.017$~\SI{}{s}, and $\Delta = 0.035$~\SI{}{s}. Figure~\ref{fig:boots_gamma_signal} shows the plot of a diffusion signal obtained with this protocol separated by shell and ordered with respect to the angle with the main fiber axis (Z-axis).
 
A bootstrapping analysis was performed to evaluate the variance of the error between the estimations with different samples sizes: $1\times 10^3$, $2\times 10^3$, $5\times 10^3$, $1\times 10^4$, $2\times 10^4$, $5\times 10^4$, $1\times 10^5$, $2\times 10^5$, $1\times 10^6$, and $2\times 10^6$ samples; and time-steps: $1 \times 10^2$, $5 \times 10^2$, $1 \times 10^3$, $5\times 10^3$, $1 \times 10^4$, and $2\times 10^4$. For each combination of the sample sized and time-steps, the signals from $50$ repetitions were generated. The error between the ground-truth and each estimated signal was computed using the Relative Mean Absolute Error (RMAE), expressed as a percentage:

\begin{equation}
RMAE(S_{gt}, S_{c}) = \frac{100}{N_g} \sum\limits^{N}_{i}{\frac{|S_{gt}(i) - S_{c}(i)|}{|S_{gt}(i)|}},
\end{equation}
where $S_{gt}$ is the ground-truth signal, $S_{c}$ is the estimated signal and $N_g$ is the number of acquisitions. The result is a total of 50 estimated error points for each sample size.



\subsection{Intra-axonal space representation}
\label{Sec:Intra}

In our second study, we look into the effect of using curved or angled geometries against straight cylinders as representations of the intra-axonal space. Such effect is of special interest on the computation of axonal diameter indexes when it is  assumed that straight cylinders capture the diffusion properties of the intra-axonal compartment. 

To understand this effect, an experiment extending the previous work from~\cite{72980:2907473} was performed, where the diffusion properties of undulating axonal substrates is studied. In our experiment, we quantified the difference on the diameter fitting estimation between parallel cylinders of constant radius and undulating cylindrical substrates.

To create curved cylindrical substrates for MCDS, a helical undulation parametrisation along $z$ was used 

\begin{equation}
U(z) = \left( A_x cos( \frac{2\pi z }{L}) , A_y sin(\frac{2\pi z}{L}), z  \right),
\label{eq:undulation}
\end{equation}

where $L$ is the wavelength and $A_x$, $A_y$ denote the amplitude in the $X$ and $Y$ axis, respectively~\cite{72980:2907473}. The amplitudes $Ax$ and $Ay$ were set to be equal to obtain helical undulations. Using the formulation above, a set of substrates was created by deforming cylinders with diameters \SI{1}{\micro\meter}, \SI{2}{\micro\meter}, and \SI{3}{\micro\meter}. The wavelength and amplitude of the undulations ranged from \SI{4}{\micro\meter} to \SI{32}{\micro\meter}  and from \SI{0.2}{\micro\meter} to \SI{2.6}{\micro\meter}, respectively; which covers a range of values of interest in the literature~\cite{72980:2907473, Haninec1986,Allison-C2004}. The resulting undulating cylindrical shapes were triangulated to use them as mesh substrates suited for MCDS. Figure~\ref{fig:undulation_samples_} presents three different substrate examples.

To compute the diameter estimation error in the intra-axonal signal, a fitting procedure was performed using an exhaustive search approach. The exhaustive search computes the RMAE between the resulting simulated signal of each undulating substrate and the analytical signal of a range of cylinders with different diameters, sampled between~\SI{0.4}{\micro\meter} and~\SI{8}{\micro\meter} with a step size of~\SI{0.01}{\micro\meter}. The analytical signals were computed using the GPD approximation for the signal in cylinders~\cite{VanGelderen1994}. The fitting procedure returns the range of plausible diameters such that the computed error between them is below a given threshold. For each undulating substrate, the threshold was fixed to a $1\%$ difference from the minimum fitting error, based on the results of the confidence study from Section~\ref{Res:Boots}.

Two different acquisition protocols were used to perform the fitting procedure. First, the original ex-vivo ActiveAx PGSE protocol~\cite{72980:2907481} explained in Section~\ref{Sec:Conf} was used. Second, we used an optimized PGSE protocol for ex-vivo axonal diameter estimation presented in~\cite{72980:2907613}. The protocol consists of a three shell acquisition with 90 orientations per shell, and a $TE=0.0359$~\SI{}{s}. The relevant parameters of each shell are as follows, i) $b=2081$~\SI{}{s/mm^2}, $G = 300$~\SI{}{mT/m}, $\delta = 0.0056$~\SI{}{s}, and $\Delta = 0.0121$~\SI{}{s}; ii) $b=3038$~\SI{}{s/mm^2}, $G = 219$~\SI{}{mT/m}, $\delta = 0.007$~\SI{}{s}, and $\Delta = 0.0204$~\SI{}{s}; and iii) $b=9542$~\SI{}{s/mm^2}, $G = 300$~\SI{}{mT/m}, $\delta = 0.0105$~\SI{}{s}, and $\Delta = 0.0169$~\SI{}{s}. Since the RMAE difference between cylinders of similar diameter depends on the protocol used an analysis of the sensitivity for each protocol was carried out.

Finally, the MC simulation parameters were chosen using a similar analysis as the one presented in Section~\ref{Sec:Conf} (not shown). The confidence estimation was computed ranging the number of particles and time steps on the substrate with higher curvature (higher amplitude and smaller wavelength) and choosing a the parameters that shows almost almost no variance on the estimations. A total of $5 \times 10^4$ particles and $5 \times 10^4$ steps were chosen to compute the signal for each individual substrate separately.


\subsection{Extra-axonal space representation}

In the case of macroscopically homogeneous substrates, e.g. with randomly packed cylinders and in absence of bundle dispersion, it has been shown that extra-axonal spins exhibit an effective diffusivity that can be described by an axi-symmetric tensor, if the volume size of the sample is high enough\unskip~\cite{Hrabe2004}. Models to estimate white matter microstructure from DW-MRI therefore assume that the extra-axonal radial contribution does not change for any direction aligned to the bundle's perpendicular plane\unskip~\cite{72980:2907595,72980:2907481,Zhang2011,72980:2907515,72980:2907503,72980:2907507,Benjamini2016}. 

Such an assumption seems to fit the validations. However, the importance of the design of the extra-axonal space has been underestimated in MCDS by assuming that substrates with any hindered configuration would match the model. To show the importance of the sample size, in terms of the number of cylinders used to construct a substrate, an analysis of the extra-axonal radial contribution in simulated signals was performed.

The radial extra-axonal DW-MRI signal was simulated for a selection of voxels with different numbers of cylinders, and a fixed distribution of diameteres and intra-axonal volume fractions. To do so, N diameters ($N=100$, $1,000$, $10,000$, $50,000$ and  $100,000$ ) were sampled from a gamma distribution with parameters $\Gamma(4.0,4.5 \times 10^{-7})$, as in our first study. The corresponding cylinders were randomly positioned in substrates with voxel size adapted such that the intra-axonal volume fraction was 60\% and ensuring periodicity at the voxel boundaries as is described in~\cite{72980:2907508}. The extra-axonal signal was simulated with the following settings: $1 \times 10^6$ particles in the extra-axonal space with diffusivity of $0.6\times 10^{-4}$~\SI{}{mm^2/s}, $TE = 0.075$~\SI{}{s}, and $1 \times 10^3$ steps. This parameters where chooseng from the previous results showed in Section~\ref{Sec:Conf}. The diffusion protocol was set to highlight the radial contribution of the diffusion signal in different diffusion time regimes as follows: $G=300$~\SI{}{mT/m}, $\delta$=0.010~\SI{}{s} and $\Delta$ from $0.015$ to $0.060$~\SI{}{s}, acquired in 180 directions evenly distributed over the xy-plane. The anisotropy of the simulated noiseless signal was quantified by computing the standard deviation of the signal divided by its mean, giving an estimate of how much the signal deviates from a perfectly radially isotropic signal.

\subsection{Framework for complex substrates generation}

Driven by the results from the previous experiments, and based on a previously published algorithm to generate tractography phantoms~\cite{72980:2907557} the following section outlines a general framework in order to generate complex substrates. We show that such framework overcomes some of the simplifications presented in the previous sections. To illustrate such capabilities, a crossing of axons bundles was generated as a study case. A qualitative evaluation was performed over the representation of crossing fibres in terms of the resulting intra-axonal volume fraction and diffusion properties in different resolutions. 

The framework is a tailored extension of the work presented in~\cite{72980:2907557}. The original framework is based in the optimization of a objective function that penalises the overlap, curvature and length of a set of initial splines called as \textit{strands}. Each strand has a constant radius used to ensure no overlapping. The optimization cost-function has the following form: 

\begin{equation}
E(\cup S) = \sum\limits_{i}^{\#S} w_o J_o(S_i) + w_c J_c(S_i) + w_l J_l(S_i),
\label{eq:FiberCostFunction}
\end{equation}
where the set $\cup S$ of size $\#S$, represents the set of all initialized strands. $S_i$ represents the strand $i$ for $i = 1,\cdots ,\#S$. The functions $ J_o(\cdot), J_c(\cdot), J_l(\cdot)$ are the overlap, curvature, and length penalization functions; and the coefficients $w_o, w_c,$ and $w_l$ are their respectively weights. Each strand $S_i$ is parametrized using a discrete set of control points that define the backbone of the strand $i$ and constant given radius; the transversal area associated with this radius is later subdivided to form sub-strands. Finally, the DW-MRI signal is then simulated by assigning a symmetric tensor along each sub-strand trajectory i.e. a simplistic local model of the micro-environment~\cite{72980:2907716}. The reader is referred to~\cite{72980:2907557} for more details. 

In our study, the aforementioned framework was modified and used to map a gamma distributed set of diameters inside the resulting strands' trajectories. The 3D-overlapping algorithm between strands implemented in the cost-function $J_o$, was also modified to make it more suitable for creating 3D meshes. This was done by computing the analytical intersection between two strands' control points, using the cylinder to cylinder collision detection described in~\cite{VANVERTH2008541}.

The result is a gamma distributed crossing configuration of deformed cylinders. The main advantage of this configuration is that the bundles inside a common area do not overlap or intersect, but interdigitate, which means that the volume is preserved in the crossing region. In addition, the curvature and length penalizations promotes a higher packing density. Finally, the proposed framework computes the DW-MRI signal by a Monte-Carlo simulation using a mesh substrate created from the configuration obtained above, instead of assigning a symmetric tensor along the sub-strands. Figure~\ref{fig:crossing_backbones} shows the crossing configuration before and after the optimization procedure.

In the presented study case, the diameters from a gamma distribution with parameters \ensuremath{\Gamma}$(1.2,1.5 \times 10^{-6})$ were sampled, resulting in a mean diameter of $\mu$ = \SI{1.8}{\micro\meter} and standard deviation $\sigma$ =\SI{1.6}{\micro\meter}, which are in the range of anatomical interest~\cite{72980:2907481}. The resulting values were truncated to avoid strands with diameters smaller than \SI{0.2}{\micro\meter}. The dimensions of the resulting enclosing volume were \SI{1200}{\micro\meter} $\times$ \SI{240}{\micro\meter} $\times$ \SI{480}{\micro\meter}; the resulting 3D geometrical crossing is shown in Figure~\ref{fig:crossing_mesh}. The 3D mesh model consists of 1,698,328 triangular faces after a post-processing of decimation and smoothing to reduce the triangle density. The total length end-to-end of the most extended strand is $1.58$ mm. The resulting diameter distribution of the overall structure is displayed in the bottom panel of Figure~\ref{fig:crossing_mesh}. 

To compute the simulated MRI signal, the total volume was divided in three voxel resolutions: $80 \times 16 \times 32$,  $40 \times 8 \times 16$, and $20 \times 4 \times 8$ voxels. A total of $105\times 10^6$ particles, and $5,000$ steps were used to compute the signal for the three resolutions. The original ActiveAx protocol~\cite{72980:2907481} from the first study was used with a diffusivity coefficient equal to $0.6 \times 10^{-3}$ mm$^2$/s  and a total diffusion time of $0.053$~\SI{}{s}.

To show qualitative results on the generated signals, the Diffusion Tensor (DT) estimation and the corresponding FA were computed using Dipy~\cite{Garyfallidis2014}, as well as the ICVF maps for each of the three resolutions. Only the shell with $b=3080$~\SI{}{s/mm^2} was used to compute the DT in each voxel. Given the lack of an analytic representation of the substrate, the ICVF was approximated by tracking the local position of the uniformly random located particles and labelling them as \textit{inside} or \textit{outside} the meshed substrate.

Finally, an evaluation of the axon diameter estimation within the crossing area was performed for the three different voxel resolutions. The axon diameter estimation was performed using the same exhaustive search method described in Section~\ref{Sec:Intra}. Only one single bundle orientation was used to compute the analytical GPD approximation; which was selected from the DT estimation at each voxel. The fitting procedure was performed using solely  the intra-axonal signal and in voxels with FA greater than $0.25$, in order to separate the effect of the extra-axonal space regarding the diameter mis-estimation.

\section{Results}

\subsection{Confidence level estimation}
\label{Res:Boots}

The overall results of the bootstrapping analysis are summarised in Figure~\ref{fig:boots_intra} and Figure~\ref{fig:bootstrapheatmap} for the intra- and extra-axonal space, and for both, the number of samples, and the number of time-steps. Figure~\ref{fig:boots_intra} shows the mean error of the 50 samples for each one of the possible combination of the selected parameters, color-coded in a heat-map. In Figure~\ref{fig:boots_intra}, we show the error of each repetition by i) fixing the number of steps to the maximum value ($2 \times 10^4 $) and varying the number of particles (left colum), and ii) fixing the number of particles to the maximum value ($2 \times 10^6$) and varying the number of steps (right column). Each data point represents one repetition of a given sample size. A total of 50 points are plotted in each row, and the mean error for each sample is highlighted with a red asterisk. The total simulation time for each repetition ranged from few seconds for the simulation with a total of $1\times 10^3$ particles to $918$ seconds for the simulation with $1 \times 10^6 $ particles. Each simulation was performed in a single node of \textit{Fidis} EPFL's cluster with 14 cores, 2.6 GHz,  and  528 MB of RAM.

For the study regarding the number of particles in the intra-axonal space, the mean RMAE between the analytical ground truth and the set of repetitions with the biggest sample size of $2 \times 10^6$ particles was of $0.47\%$. For the extra-axonal space, the mean RMAE between the computed gold-standard with $20 \times 10^6$ and the set with $1 \times 10^6$ particles was equal to $0.71\%$.  For the analysis varying the number of time-steps, the minimum mean RMAE achieved was of $0.44\%$ for the intra-axonal space and $0.38\%$ for the extra-axonal. The difference between the mean RMAE between $5 \times 10^3$ and $2 \times 10^4$ was less than $0.2\%$ for both the intra- and extra-axonal space.

\subsection{Intra-axonal space representation}


The range of diameters, computed from our fitting procedure on both protocols, are displayed in Figure~\ref{fig:undulation_tables}. Each cell is coloured according to its minimum RMAE. An amplitude (amp) of \SI{0}{\micro\meter} corresponds to a straight cylinder which presented the minimum fitting error achievable for each diameter. Values with the highest amplitude and lowest wavelength (wl) corresponds to the axons with the highest undulation ( amp = \SI{2.6}{\micro\meter}, wl = \SI{4}{\micro\meter} ); on the other hand, values with the lowest amplitude and highest wavelength ( amp = \SI{0.2}{\micro\meter}, wl = \SI{32}{\micro\meter}) corresponds to almost straight axons.

The protocols' sensitivity analysis is shown in Figure~\ref{fig:cylinders_cross_MAE} which presents a visualisation of the RMAE between the analytical signal of straight cylinders with different diameters. Regions with homogeneous values are difficult to differentiate between each other, e.g. the region with diameters between \SI{0}{\micro\meter} and \SI{2.0}{\micro\meter}. The coloured line shown in both plots marks the $1\%$ level curve. In this plot, the protocols' contrast in function of the cylinder's diameter can be visualized, which correlates with the intervals showed in Figure~\ref{fig:undulation_tables}.

\subsection{Extra-axonal space representation}

Three different substrates (with $100$, $1,000$ and $10,000$ cylinders, corresponding to voxel sizes of \SI{23x23}{\micro\meter}, \SI{71x71}{\micro\meter}, and \SI{230x230}{\micro\meter}, respectively) and their corresponding radial DW-MRI signals, are shown in Figure~\ref{fig:effect_of_size_substrates}. The shown voxel sizes were chosen to highlight the radial anisotropy in three representative sizes. The substrate with $10,000$ cylinders, i.e. with the biggest voxel size, had the most isotropic radial DW-MRI signal. On the other hand, the most anisotropic signal was observed for the substrate with the smallest number of cylinders. Figure~\ref{fig:effect_of_size_stats} shows the mean and standard deviation of the radial extra-axonal signal as a function of the voxel size. The same experiment (not shown) was conducted using cylinders with higher diameter. Results indicated that the number of cylinders was the limiting factor. Indeed, the mean of the radial extra-axonal signal also converged for 10,000 cylinders, but this time a voxel size of approximately~\SI{400x400}{\micro\meter} was required to generate isotropic profiles instead of the \SI{230x230}{\micro\meter} limit observed for a distribution with smaller cylinders.

\subsection{Framework for complex substrates generation}

The resulting crossing with two fibre populations is outlined in Figure~\ref{fig:crossing_mesh}. The total optimization time to create the substrate was around 42 hours, where most of the optimization time (about 35 hours) was needed in the second optimization iteration, after the subdivision on gamma distributed radii, that ensure that no small overlaps were introduced due to the subdivision and abrupt angular changes. The optimization was performed using a single core 2.8 GHz CPU. On the other hand, the total simulation time for the full geometry with $105\times 10^6$ particles was less than 24 hours using a total of 8 nodes with 28 cores on \textit{fidis} EPFL's cluster with 6GB of RAM per node (48GB in total).

The resulting diffusion tensor and FA maps are shown in Figure~\ref{fig:crossing_DT_FA_maps} for the three different resolutions. Local diffusivity changes, as well as signal alterations related to the curvature of the individual axons, can be observed. 


Figure~\ref{fig:ICVF_maps} shows the intra-axonal volume fraction in all resolutions. In the highest resolution, small water compartments can be seen in the crossing sections; this is an effect of the optimisation procedure which ensures no overlapping fibres. In the lowest resolution, such compartments are no longer visible, but they are reflected in the decrease of the intra-axonal volume fractions.  

Figure~\ref{fig:diameter_maps} shows a visualization of one plane of the axon diameter estimation maps of the volumetric region highlighted in Figure~\ref{fig:ICVF_maps}, and the obtained diameter distribution for the three resolutions. The higher resolution ($80 \times 16 \times 32$) estimation includes a total of 2848 voxels, while the lowest resolution contains a total of 112 voxels. Figure~\ref{fig:crossing_mesh} bottom-right panel shows the resulting sampled diameters inside the crossing configuration, which is noticeably skewed to smaller diameters; this is an effect of the packing algorithm inside individual circular strands which under-represent the tail of the distribution because of the difficulty of packing strands with big diameters. This effect will irremediably affect the effective apparent radius $r_{eff} \approx (\frac{<r^6>}{<r^2>})^{1/4} $ ~\cite{72980:2907510} given by the intra-axonal contribution of the signal. The resulting effective diameter of the conjoint assemble of strands was $2*r_{eff}$ = \SI{3.48}{\micro\meter}, which in average agrees with the estimated mean diameters shown in Figure~\ref{fig:diameter_maps} on each resolution.

\section{Discussion}

In the past two decades, the research community has used MCDS to generate and validate MR diffusion data and microstructure models~\cite{72980:2907478,72980:2907473,72980:2907499,72980:2907552,72980:2907611,72980:2907508,72980:2907480,72980:2907477}. However, questions have been raised on the accuracy of the simplified geometries used to create the diffusion substrates\unskip~\cite{72980:2907474,72980:2907480,72980:2907473,72980:2907485}, emphasising the need of highly-validated and reproducible simulations. Such oversimplifications have been proven not to capture the complexity of the axonal structures of white matter, and thus its diffusion characteristics\unskip~\cite{72980:2907485}. Moreover, it can be argued that the use of such elementary geometries—used as backbone in the microstructure models—as a ground-truth, not only introduce a systematic bias that inherently supports the evaluated method, but also misapplies the very purpose of using Monte-Carlo simulations. In this work, we outlined pitfalls encountered in the design of such simulations. Our experiments showed how the design of each substrate compartment is likely to introduce an estimation bias if it is not addressed appropriately. 

Our first study specifically shows the effect of an inappropriate selection of parameters on the reproducibility of a estimated signal, which could also skew an analysis toward inaccurate results. Differently to previous studies~\cite{72980:2907508}, we compute the extra-axonal ground-truth from high-quality simulations, avoiding the use of tortuosity models that could introduce a bias because of their oversimplifications. The error in the estimation presented in Figure~\ref{fig:boots_intra} illustrates the great amount of possible estimation variability for a relatively simple substrate. We found that the signal on each compartment showed a high variability for simulations with less than $5 \times 10^5$ particles and $1 \times 10^4$  steps. We can extrapolate from this that any estimation from more complicated substrates, such as the ones with undulation or crossings, or even higher diffusivity, will likely entail even higher variability. In order to avoid such uncertainty on the estimations for more complicated substrates a similar analysis as the one presented should be procured. 

In our second study, we explored the effect of breaking the assumption of straight cylinders as the intra-axonal representation in function of the apparent diameter estimation. The helical representation used in this study, while reported to appear in the nervous system, maybe not be an accurate representation of the axonal angular variations along the longitudinal direction in the brain white matter, specially in the micro-scale. However, it gives us a convenient starting point to study the effect of angular variations in the intra-axonal compartment over the diffusion signal, a theoretical analysis on this type of structures can be found as well in~\cite{JanSamo} (in press). From this study, we found a considerable mis-estimation in the presence of undulation for both protocols and in the three studied diameters. The relative fitting error for the smaller diameter (\SI{1}{\micro\meter}) was the higher among the three cases (more than $300 \%$ for some cases). Previously,~\cite{Nilsson2017} proposed a formulation to compute the resolution limit for parallel cylinders using standard single-shell PGSE sequences. According to this formalism, the minimum differentiable diameter is $d_{min} = \frac{768}{7} \sigma D(\gamma^2 \delta |G|^2)^{-1}$, where $\sigma$ is the significance level, defined as the minimum tolerated percentage of signal change. For a fixed value $\sigma = 1\%$ change, the resolution limit predicted for both protocols used in this study were $d_{min} =$\SI{1.28}{\micro\meter} for $|G|=140 mT/m$, and $d_{min} =$\SI{1.02}{\micro\meter} for $|G|=300 mT/m$. However, such estimates are based on a number of assumptions which does not hold in our experimental conditions. For example, the formulation is valid for parallel and straight cylinders and for acquisition protocols with a single shell with parameters $\delta = \Delta$. As in this experiment we are studying non-parallel and curved cylinders with multi-shell protocols with $\Delta >> \delta$, we performed a numerical sensitivity analysis to obtain more accurate resolution limits. From the resulting plots showed in Figure~\ref{fig:cylinders_cross_MAE}, it can be seen that the signal originated from cylinders with diameters below \SI{2.5}{\micro\meter} for the first protocol, and \SI{2.0}{\micro\meter} for the second, are virtually indistinguishable. On the other hand, diameters above \SI{3.0}{\micro\meter} have more significant RMAE, which make them easier to differentiate. We also observed that the range of diameters from our fitting method did not follow a simple trend between protocols; that is to say, increments on the undulation parameters, which effect can be summarized in terms of the  tortuosity factor~\cite{72980:2907473} $\lambda = \sqrt{ (2\pi A /  L)^2 + 1}$, does not follow a simple relationship between protocols (horizontal axis of the results in Figure~\ref{fig:undulation_tables}). This is likely to be an effect of the parameters of the acquisition protocol ($\delta$, $\Delta$, and the $TE$), which vary between shells and thus changing the effective diffusion time. From a comparison of both protocols, we corroborated that the optimized protocol showed better results in terms of the fitted diameter and range of similar diameters. However, there was still a considerable mis-estimation, especially for the undulation of  \SI{1}{\micro\meter} diameters. We consider this experiment to be of great interest for any future protocol optimization or diameter estimation framework, since it illustrates how sensible the estimation of the axon's diameter based on the cylindrical model are, even for regular and smooth angular deviations.


Our third experiment showed that a sufficiently rich sampling is required for the simulated signal to converge. Indeed, small substrates have a limited number of cylinders, limiting the variability of hindered micro-environments sampled by the spins during the M.C. simulation---yielding anisotropic patterns in the radial DW-MRI signal. The results also showed a bias in the mean amplitude, with small voxels having lower signal than bigger voxels. Our results suggest that, for a given diameter distribution, substrates with an area smaller than \SI{200x200}{\micro\meter} will present biased extra-axonal signals. Such results are in accordance with previously reported results~\cite{72980:2907565} in terms of the voxels’ size. However, this lower bound probably depends on the distribution of diameters and cylinder packing on one side, as well as the typical diffusion length of the spins, given by their diffusivity and the diffusion time of the experiment.

Finally, as part of our effort to create more  realistic substrates, we outlined a framework to tackle the challenging problem of creating non-overlapping crossing configurations that preserves the volume fractions between the non-crossing and crossing area, while enforcing a high packing density. Configurations which mimic better real tissue~\cite{Biomecanic_book}, are important since they provide a more challenging environment to test and validate microstructure models and even tractography methods, in contrast with naive crossings which have been proven to be indistinguishable from a simple superposition of individual fascicles~\cite{Gaythan}. From the diffusion tensor and FA maps shown in Figure~\ref{fig:crossing_DT_FA_maps} we can observe the presence of multiple compartments as an effect of the volume preservation condition. Also, Figure~\ref{fig:ICVF_maps} shows how the intra-axonal volume fraction changes as the resolution decreases. Such information can be used to study the microstructure information in the presence of several diffusion compartments and volume fractions in different resolutions without using an explicit interpolation. This decrease of the ICVF is an effect of the presence of dispersion and deformation of the fibre bundles. However, even in the lowest resolution, the intra-axonal volume fraction achieved was over $48\%$, which is considerably higher than the icvf (of $20\%$) of a previously presented framework for generating realistic numerical phantoms for crossing fascicles~\cite{Ginsburger2018}. By optimizing the penalization term of the strands' curvature in our framework, we expect to be able to achieve even higher packing densities---closer to the expected ones from the brain's white matter tissue. On the other hand, the diameter estimations computed over the merging area of the two fibre populations, showed a overestimation in accordance with the results of Section~\ref{Sec:Intra}. Such mis-estimation can be explained by angular perturbations in the fibre trajectory in both the micro- and meso- scale of the simulated fibres. In previous studies, the axon diameters were overestimated by factors ~3–5 in clinical scanners (Alexander et al., 2010; Zhang et al., 2011). This bias was attributed to the insensitivity of the measurement schemes to small axons (Dyrby et al., 2012), the noise, or the commonly neglected time-dependence of diffusion in the extra-axonal space~\cite{DeSantis2016a}. The diameters reported in this study were estimated by using only the intra-axonal signals, thus the overestimation can be explained only by the dMRI signal insensitivity to the smaller axons and by the signal changes due to axon undulations and microscopic dispersion. This renders our estimations as a best case scenario. 
 
\subsection{Considerations and future work}

The generalisability of the results presented above is subject to certain limitations. For instance, in-vivo diffusion and protocol settings, the use of non-regular deformations in the intra-axonal substrates, and the joint study of the intra- and extra- axonal space, may affect the results toward higher variability or mis-estimations of the axon diameters. In addition, a number of structural features present in white matter tissue —such as the axonal myelin sheath, Ranvier nodes, or diameter changes along the axons trajectory—are missing. Because of this, the results presented above should be taken as a type of lower bound in terms of the minimum parameters needed (for the number of samples and time-steps) and possible mis-estimations (in terms of our axon diameters estimates). 


Notwithstanding these limitations, we consider that the aforementioned framework, complemented with the optimized simulator developed, are able to overcome the simulations pitfalls presented in this work. In addition, the parameter selection analysis presented in this work provides a way to ensure the reproducibly of the Monte-Carlo simulations. A thorough study of the properties of more complex substrates generated with the proposed framework is beyond the scope of this study. Future research should therefore concentrate on the generation and study of such configurations, which may help the DW-MRI research community to generate more reliable ground-truth data.

\section{Conclusions}

The main contribution of this work can be summarized in three main aspects. First, this paper outlines and investigates a set of pitfalls encountered on the parameter selection and substrates' design for Monte-Carlo simulations. Our results over the effect of the number of particles and time-steps, as well as our quantification over the effect of the substrate's size on the extra-axonal space can be immediately taken to evaluate the design of future experiments. In overall, we found that for experiments with parameters in the range used in this study---which are in the range of interest in the literature---simulations with less than $5 \times 10^5$ particles and $1 \times 10^4$  steps carried a significant variance between the computed signals for both, the intra- and extra-axonal compartments. In addition, we found that simulations substrates with less than 10,000 sampled cylinders induced an important bias on the directional symmetry of the diffusion signal in directions transversal to the the main fiber direction. Such parameters are almost one order of magnitude bigger than previously used on the literature, which inherently affects the reproducibility of such results~\cite{RENSONNET2,Gaythan,72980:2907481,72980:2907508}. Second, our evaluation over the effect of introducing angular perturbations in the intra-axonal space representation---by means of the estimated axon diameter based on the cylindrical model---showed a considerable deviation from the expected results. This results are somehow in agreement with previous findings and contributes additional evidence that suggests that performing whole brain axon diameter estimation is still far from being straightforward using simplified models, such as the straight cylindrical diffusion model. Finally, this paper presents a framework able to generate complex fibre configurations with desired microstructure information based on a previous algorithm used to create tractography phantoms. We showed the framework's capabilities to generate complex fibres configurations which, along with the simulator developed in this work, are able to generate more challenging and composite Monte-Carlo simulations.

We consider that the results presented in this work, along with the reported procedure to evaluate the estimations' variability, the substrate generation framework, and the simulator developed, pave the way towards more realistic and reproducible Monte-Carlo simulations for Diffusion-Weighted MRI.

\begin{appendices}



\section{Simulation Complexity}
\label{App:B}

\begin{algorithm}
\caption{MCDS core algorithm}\label{alg:MCDS}
\begin{algorithmic}[1]
\Procedure{BasicSimulation}{}
	\State $\textit{$N_s$} \gets \text{Number of }\textit{spins}$
	\State $\textit{$N_t$} \gets \text{Number of }\textit{time-steps}$
	\State $\textit{$N_g$} \gets \text{Number of }\textit{acquisitions}$

	\For{each $N_s$ spins}:
		\For{each $N_t$ time step}:
			\State $UpdateSpinPosition(\dots)$
					
		\EndFor
		
		\For{each $N_g$ acquisition}:
			\State $UpdateTotalDephase(\dots)$
		\EndFor		
	\EndFor
	
	\For{each $N_g$ acquisitions}:
		\State $ComputeDWSignal(\dots)$
	\EndFor

\EndProcedure
\end{algorithmic}

\end{algorithm}


Algorithm~\ref{alg:MCDS} shows the fundamental parts of a diffusion simulation. In~\cite{72980:2907508} they summarized the \textit{complexity} of a simulation as $U = N_t*N_s$ i.e. the number of steps times ($N_t$) times the number of simulated particles ($N_s$). From an experimental point of view, such a formula is useful to summarize the relation between the number of samples and the temporal resolution for the quality of the estimation, however, the computational complexity is not accurately characterized. This because such complexity takes into account only the first two nested loops and intrinsically assumes that the position update has complexity O(1), which is only true in the absence of a substrate, i.e. free diffusion. In here, we proposed the following asymptotic complexity formula: $\mathcal{O}(N_s*N_t*N_o) + \mathcal{O}(N_s*N_g)$; which incorporates the effect of the collision detection (in terms of the number of simulated obstacles $N_o$)  and number of acquisitions ($N_g$). From this expression, it can be seen that the number of particles N\ensuremath{_{s}} linearly increases the computational burden of a simulation. At the same time, the quality of the estimated signal will greatly depend on the amount of sampled particles. \cite{72980:2907508} and~\cite{72980:2907474} both performed an analysis of the effect of the number of particles for simple geometries, showing how intricate can be the selection of the number of samples.\\

\begin{algorithm}
\caption{Particle Position Update}\label{alg:updatePos}
\begin{algorithmic}[1]
\Procedure{UpdateSpinPosition}{}
	\State $\textit{$\Omega$} \gets \text{List of }\textit{Obstacles}$
	
	\Do
		\For{each obstacle in $\Omega$}
			\State checkForCollision($\dots$)
			\If {collision detected}
				\State reflectTrajectory($\dots$)
			\EndIf
		\EndFor			
	\doWhile{collision is detected}
							
\EndProcedure
\end{algorithmic}

\end{algorithm}

Algorithm~\ref{alg:updatePos} shows the basic operation of each step update. To address the complexity of the overall function, the expected amount of collision has to be computed. However, the expected amount of collision will depend on the step-length, the number of obstacles, the separation between them, and diffusion parameters; making it impractical to compute. On the other hand, the method \textit{checkForCollision()} has $\mathcal{O}(N_o)$, where $N_o$ is the total number of obstacles (triangles, cylinders, spheres, etc.) in the substrate. Spatial optimization procedures, such Axis Aligned Bounding Boxes (AABB) or R-Trees, can optimize the collision detection by splitting and search the obstacle domain in $\Omega(log(N_O))$ for well balanced spatial structures~\cite{Agarwal:2001}. By neglecting the complexity of the multiple reflection of a single step, we can summarize the complexity of the first two nested loops in Algorithm~\ref{alg:MCDS} as $\mathcal{O}(N_s*N_t*N_o)$. Finally, since updating the total dephasing in each iteration can be done in constant time, the second nested loop has complexity $\mathcal{O}(N_s*N_g)$, where $N_g$ is the number of acquisitions (number of output signals). The overall simulation complexity is then rendered as: 
	
\begin{equation}
\mathcal{O}(N_s*N_t*N_o) + \mathcal{O}(N_s*N_g) = \mathcal{O}\big(\max(N_s*N_t*N_o,N_s*N_g)\big) 
\label{eq:complexity}
\end{equation}

The equation above summarizes how sensitive a simulation is to the simulation parameters. The number of particles $N_s$ has the biggest impact since it scales the complexity in both terms, in addition, it's usually the biggest one followed by the number of steps $N_t$. On the other hand, the number of acquisitions or output signals $N_g$ is usually the smaller one and therefore the complexity above can be usually reduced as $\mathcal{O}\big(N_s*N_t*N_o\big)$. Nevertheless, we cannot neglect it from the complexity above, since it may have a bigger impact in application related to protocol optimization or q-space exploration, where a big number of shells or sampling direction is needed. Finally, the number of obstacles $N_o$ may variate from hundreds~\cite{72980:2907481} to millions, as in our presented framework.

\end{appendices}

\section*{Acknowledgements}

We would like to acknowledge Dr Gary Zhang for his advice that highly contributed to the development of this work. We also thank Thomas Yu for his generous help proofreading this manuscript.

This project has received funding from the Swiss National Science Foundation under grant number 205320\_175974.

We gratefully acknowledge the support of NVIDIA Corporation with the donation of the Titan Xp GPU used for this research.

This work was supported by EPFL through the use of the facilities of its Scientific IT and Application Support Center.

This project has received funding from the European Union's Horizon 2020 Framework Programme for Research and Innovation under grant agreement 665667 (call 2015) and from the Natural Sciences and Engineering Research Council of Canada (NSERC).

This work is supported by the Center for Biomedical Imaging (CIBM) of the Geneva-Lausanne Universities and the EPFL, as well as the foundations Leenaards and Louis-Jeantet.\\


Conflict of Interest: None

\bibliographystyle{abbrv}
\bibliography{main}
\newpage

\begin{figure}[h]
\centering
\includegraphics[width=120mm]{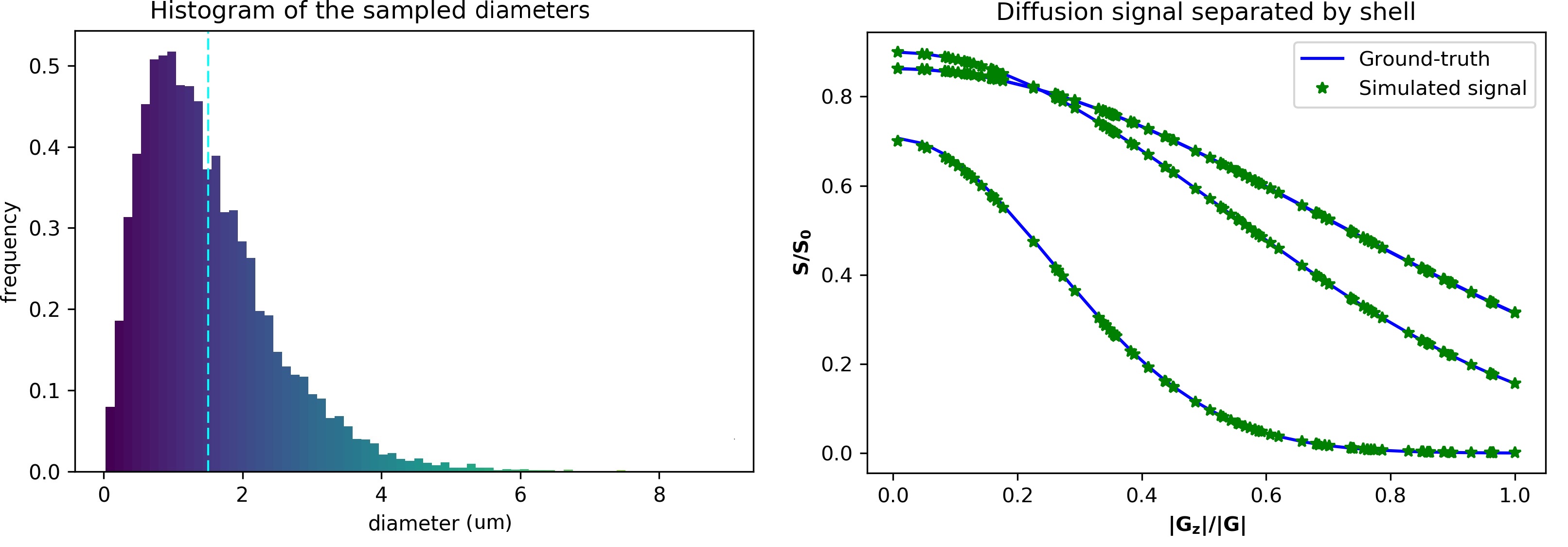} 
\caption{Gamma distributed radii and corresponding intra-axonal diffusion signal. Left panel: The distribution of the sampled diameters, the dotted line marks the sampled distribution mean. Right panel: The computed ground-truth along with the simulated signal used for the intra-axonal space representation. A total of four curves are plotted corresponding to each b-value =  \SI{1925}{s/ mm^2}, \SI{1932}{s/ mm^2}, \SI{3093}{s/ mm^2}, and \SI{13191}{s/ mm^2}. The curves corresponding to a b-value = \SI{1925}{s/ mm^2} and \SI{1932}{s/ mm^2} are completely overlapped and corresponds to  the lowest decay. The signals of each shell are ordered by the normalized Z coefficient of the gradient direction.}
\label{fig:boots_gamma_signal}
\end{figure}

\begin{figure}[h]
\centering
\includegraphics[width=85mm]{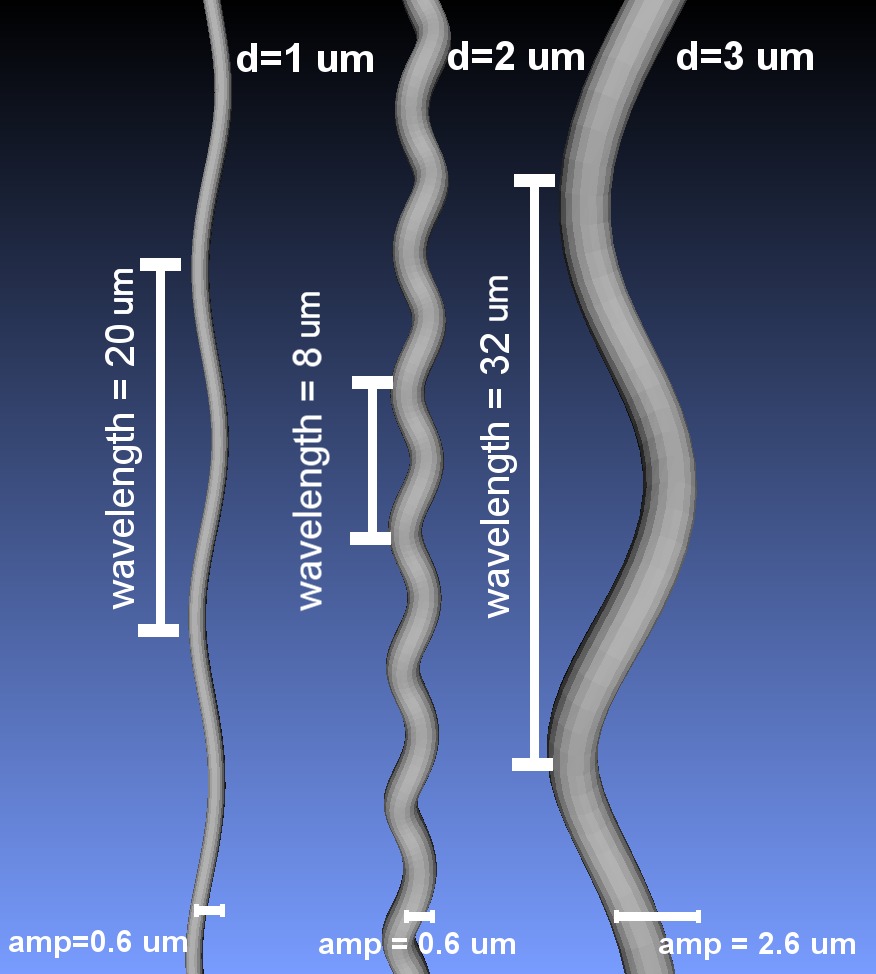}
\caption{Examples of the curved meshes used as intra-axonal substrates in this study, for three different diameters and different undulation parameters.}
\label{fig:undulation_samples_}
\end{figure}

\begin{figure}[hbtp]
\centering
\includegraphics[width=85mm]{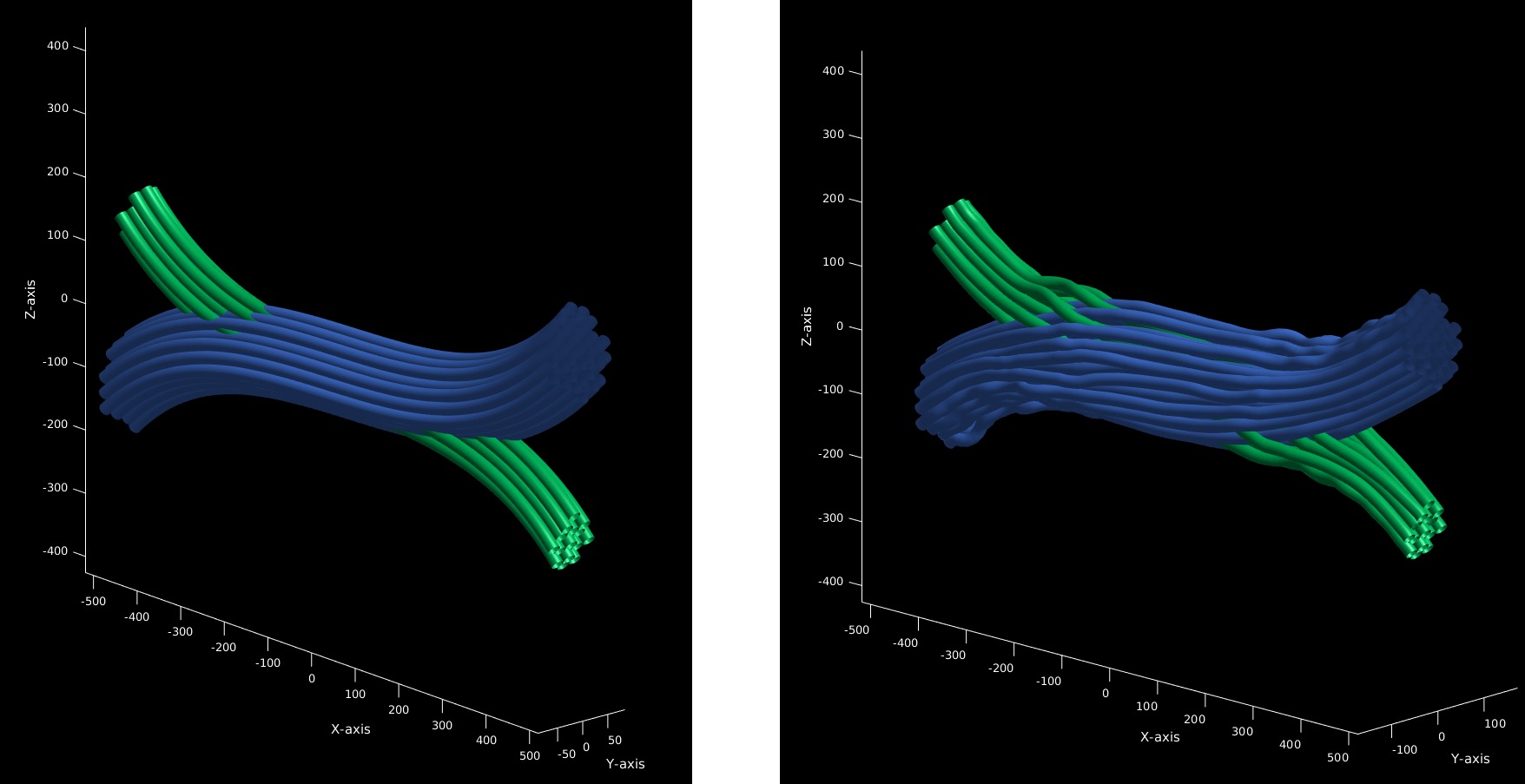}
\caption{Optimisation procedure of initial trajectories. Left panel: initial trajectories parametrised as a set of control points with constant radius. Right panel: the resulting trajectories after the optimisation procedure which ensures that there is no overlapping between the resulting strands.}
\label{fig:crossing_backbones}
\end{figure}

\begin{figure}[h]
\centering
\includegraphics[width=85mm]{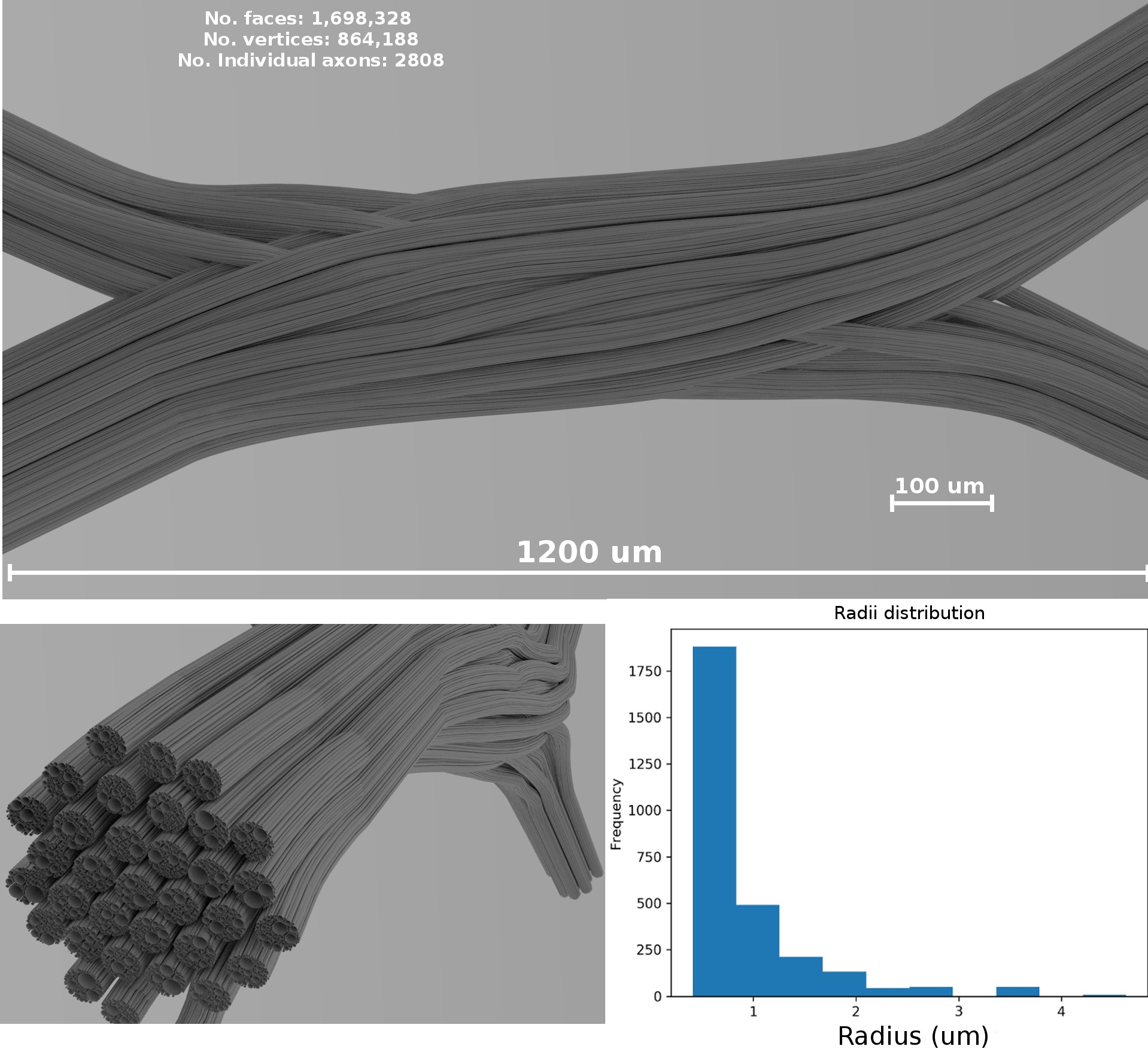} 
\caption{ Top panel shows a visualisation of the resulting fibre crossing substrate after the strand refinement and the smoothing and decimation of the triangular faces. Left-bottom panel shows the resulting sub-strand configuration of one of the crossings bundles. Right-bottom panel shows the overall diameter distribution of the displayed bundle on the left. A rendered video of the full crossing is included as supplementary material.}
\label{fig:crossing_mesh}
\end{figure}

\begin{figure}[h]
\centering \makeatletter\IfFileExists{Figures/Joint_bootstrapping.jpg}{\includegraphics[width=180mm]{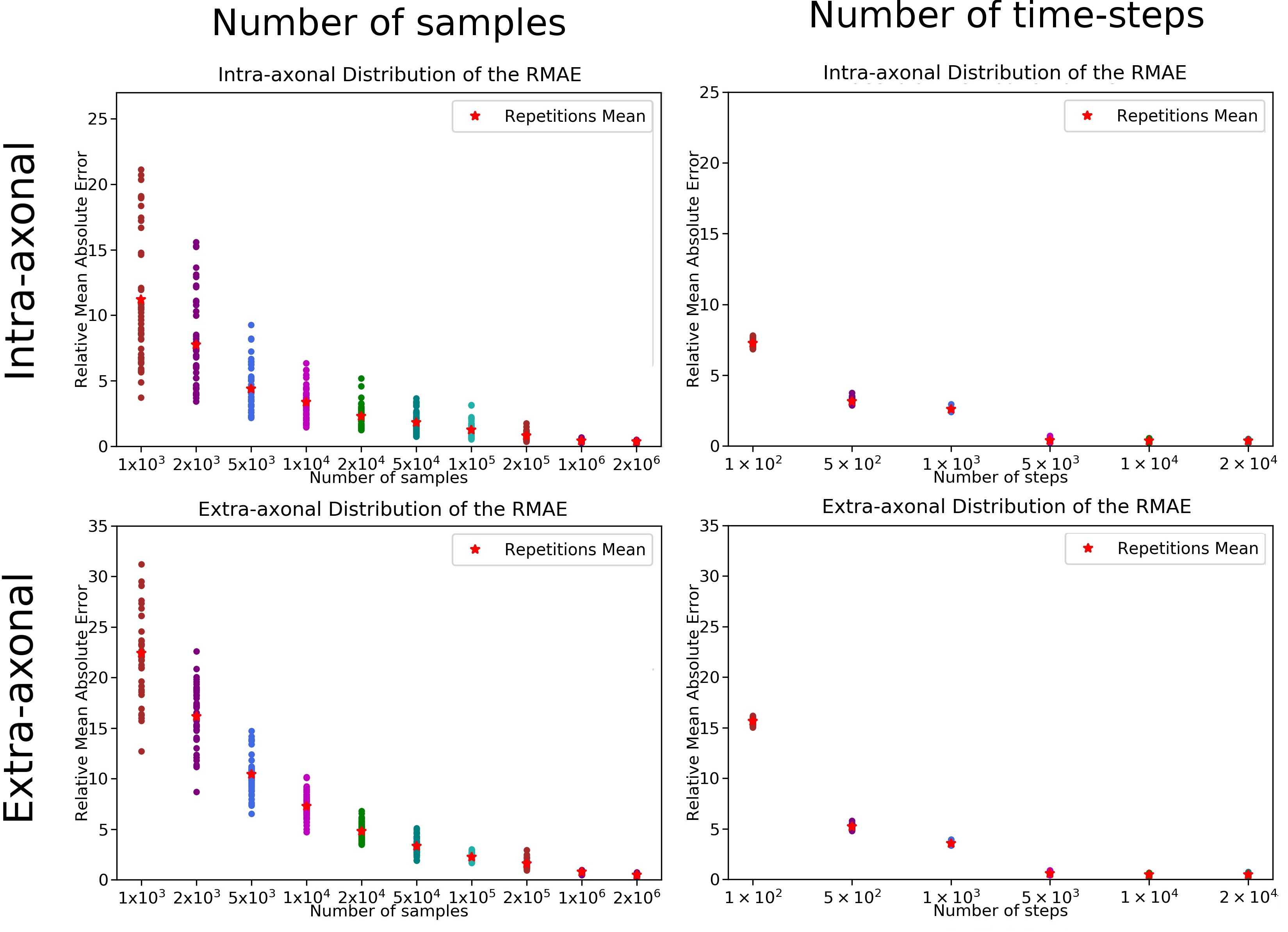}}{}
\makeatother 
\caption{RMAE for each repetition and sampled size for (left) the number of samples and (right) number of time-steps. The two panels on the top row correspond to the intra-axonal results, and the bottom row to the extra-axonal. The $X$-axis shows each sample size, and the Y-Axis shows the RMAE of all the repetition in same colour. The mean RMAE of all the repetitions is depicted with a red marker.}
\label{fig:boots_intra}
\end{figure}

\begin{figure}[h]
\centering
 \textbf{Mean RMAE error}\par\medskip
\minipage{0.4\textwidth}
\centering
 \textbf{Intra-axonal}\par\medskip
  \includegraphics[width=\linewidth]{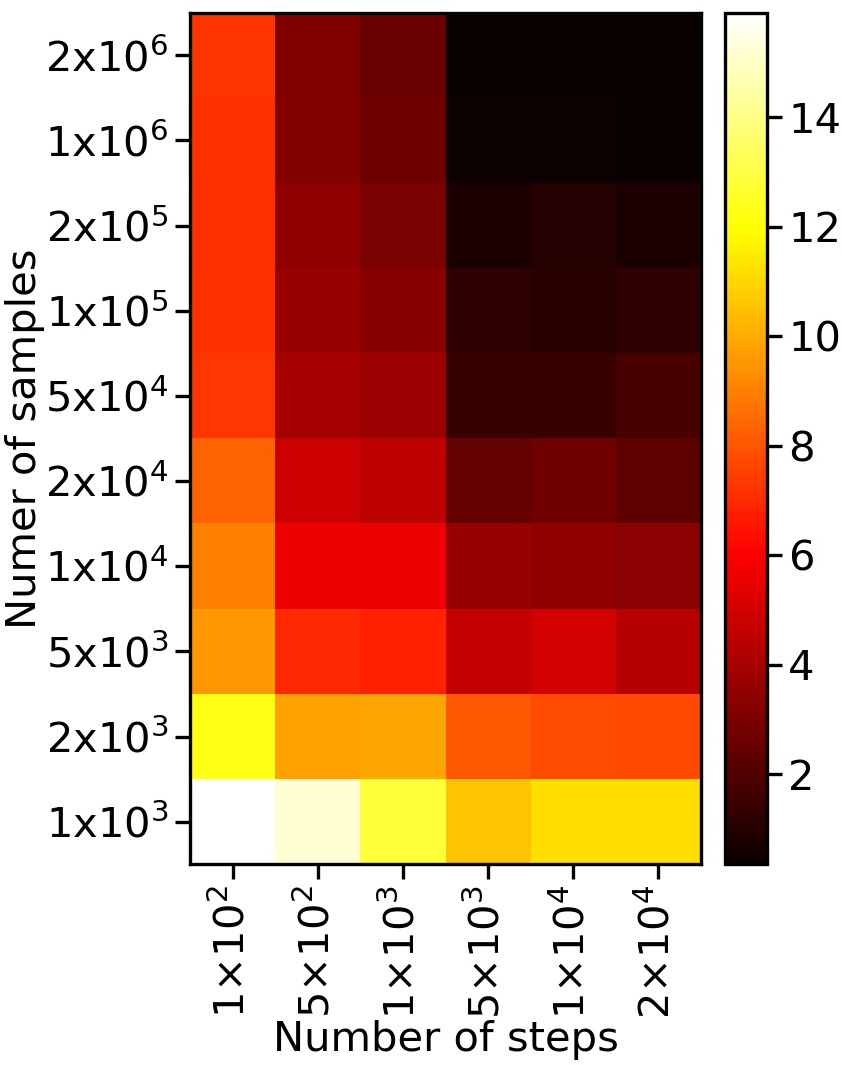}    
\endminipage\hspace*{1cm}
\minipage{0.4\textwidth}
\centering
 \textbf{Extra-axonal }\par\medskip
  \includegraphics[width=\linewidth]{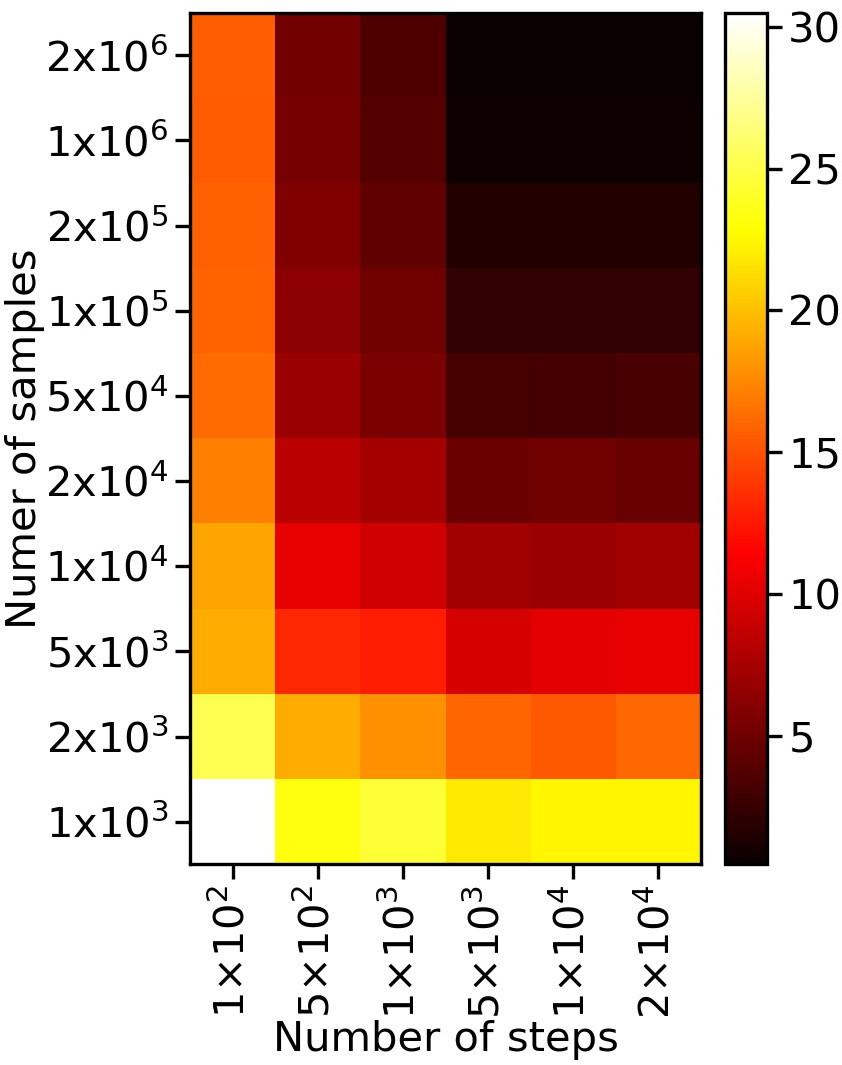}
\endminipage

\caption{ Heat map of the mean RMAE for all the combinations between the number of steps and the number of samples. Each cell corresponds to the mean RMAE of the 50 repetitions.}
\label{fig:bootstrapheatmap}
\end{figure}

\begin{figure}[h]
\centering \makeatletter{\includegraphics[width=180mm]{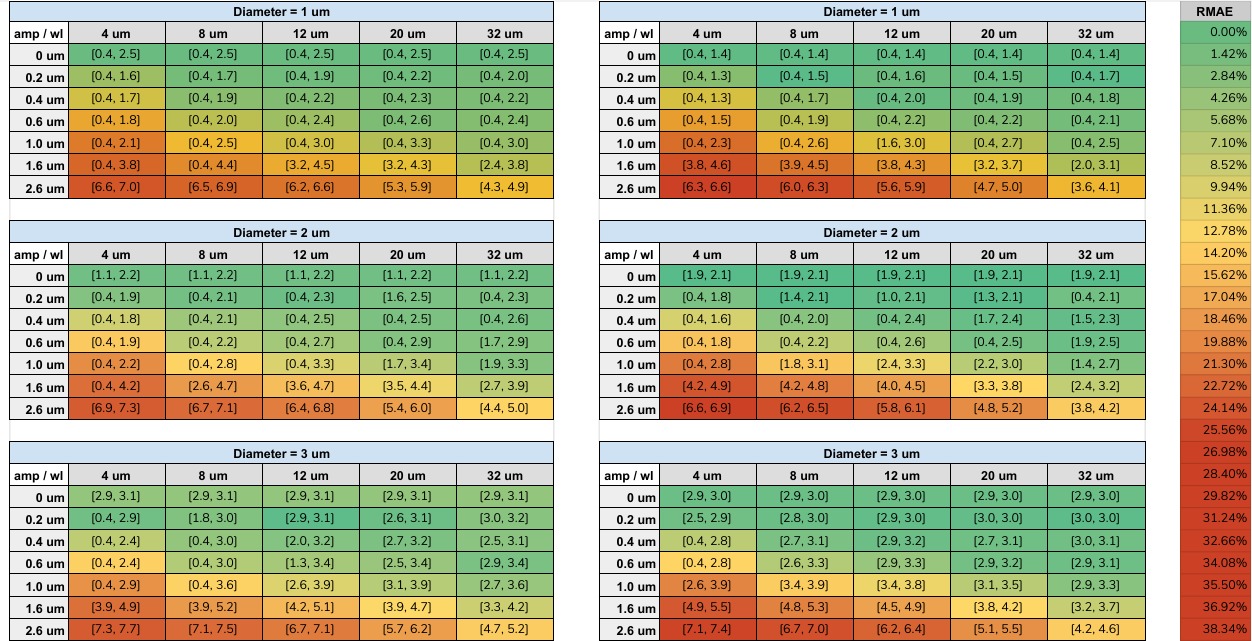}}
\makeatother 
\caption{ Tables of the fitting results. Left column shows the fitted intervals of the original ex-vivo ActiveAx protocol~\cite{72980:2907481} and the right column of the optimized ex-vivo protocol from~\cite{72980:2907613}, for the three simulated diameters. The min and max diameters (\SI{}{\micro\meter}) of the fitted range are listed between the square brackets for each simulated amplitude and wavelength. The colour of each cell is encoded with respect the minimum RMAE in the fitted range accordingly to the colour-bar on the right.}
\label{fig:undulation_tables}
\end{figure}

\begin{figure}[hbtp]
\centering
\includegraphics[width=85mm]{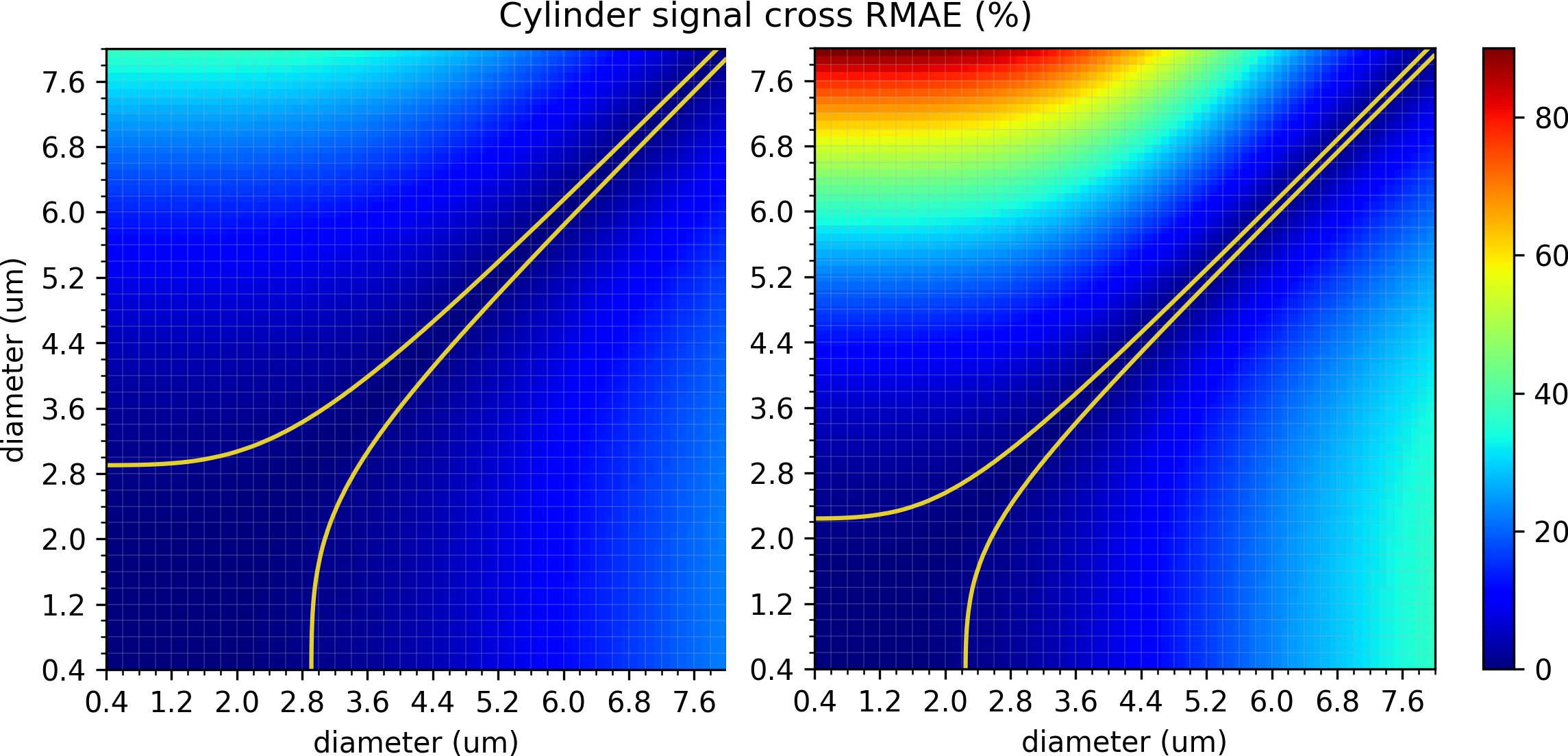} 
\caption{The in-between RMAE of the analytical signal of a cylinder, obtained using the GPD approximation, for the range of diameters used in this study. The original ex-vivo ActiveAx acquisition protocol~\cite{72980:2907481} is displayed on the left panel, and the optimized ex-vivo acquisition protocol from~\cite{72980:2907613} on the right panel. Values of the diagonal correspond to the RMAE of two straight cylinders of the same diameter and therefore equals to 0. The coloured line shown in both plots marks the $1\%$ difference level curve.}
\label{fig:cylinders_cross_MAE}
\end{figure}

\begin{figure}[h]
\centering \makeatletter\IfFileExists{Figures/effect_of_size_substrates.png}{\includegraphics[width=180mm]{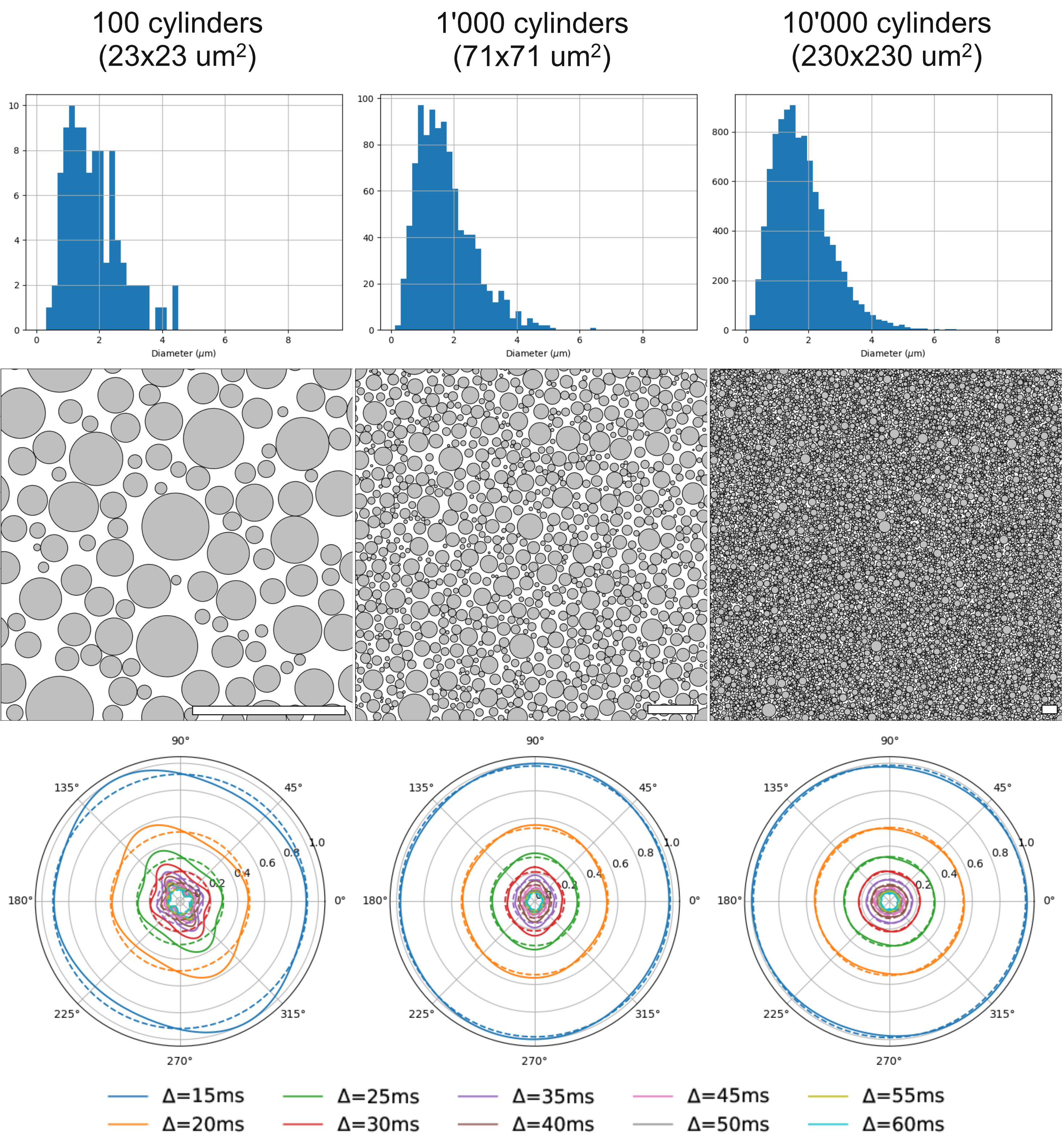}}{}
\makeatother 
\caption{Results for 3 substrates, with 100, 1,000 and 10,000 cylinders respectively. First row: sampled diameter distributions for each voxel-size, getting closer to the desired distribution law as the voxel-size increases. Second row: cylinder positions in each substrate. White scale bar corresponds to 10 $\mu$m. Third row: radial DW-MRI signal simulated from the respective substrates. Each coloured line corresponds to one different $\Delta$ duration. Dotted lines correspond to the mean radial signal for each diffusion time.}
\label{fig:effect_of_size_substrates}
\end{figure}

\begin{figure}[!htbp]
\centering \makeatletter\IfFileExists{Figures/effect_of_size_stats.png}{\includegraphics[width=180mm]{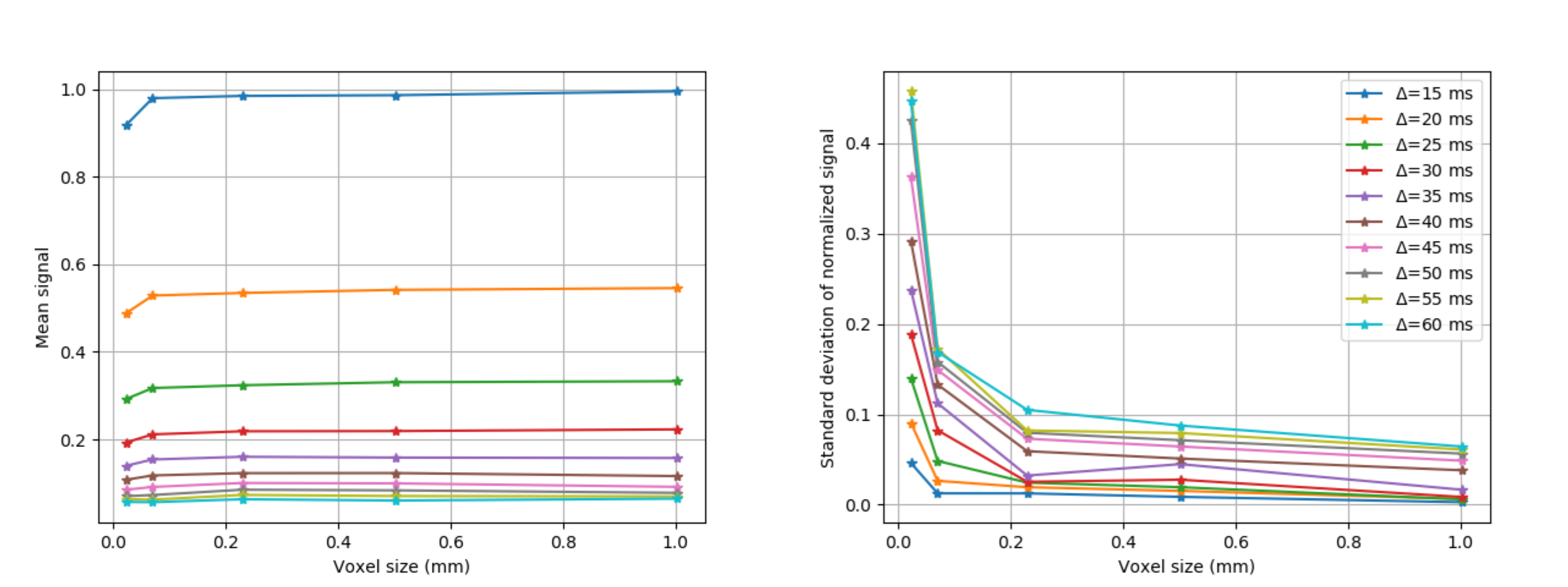}}{}
\makeatother 
\caption{{Mean and standard deviation of the radial DW-MRI signal as a function of substrate size. The signal is shown for each of the different $\Delta$.}}
\label{fig:effect_of_size_stats}
\end{figure}

\begin{figure}[h]
\centering \makeatletter\IfFileExists{DT_FA_maps.jpg}{\includegraphics[width=180mm]{DT_FA_maps.jpg}}{}
\makeatother 
\caption{ From the leftmost to the right: diffusion tensor map, the resulting fractional anisotropy and the two highlighted ROIs in each map respectively. Each image corresponds to the same volume slice in the $XZ$-plane. The ROI's highlights one area where different compartments result from the optimization procedure.}
\label{fig:crossing_DT_FA_maps}
\end{figure}

\begin{figure}[hbtp]
\centering
\includegraphics[width=85mm]{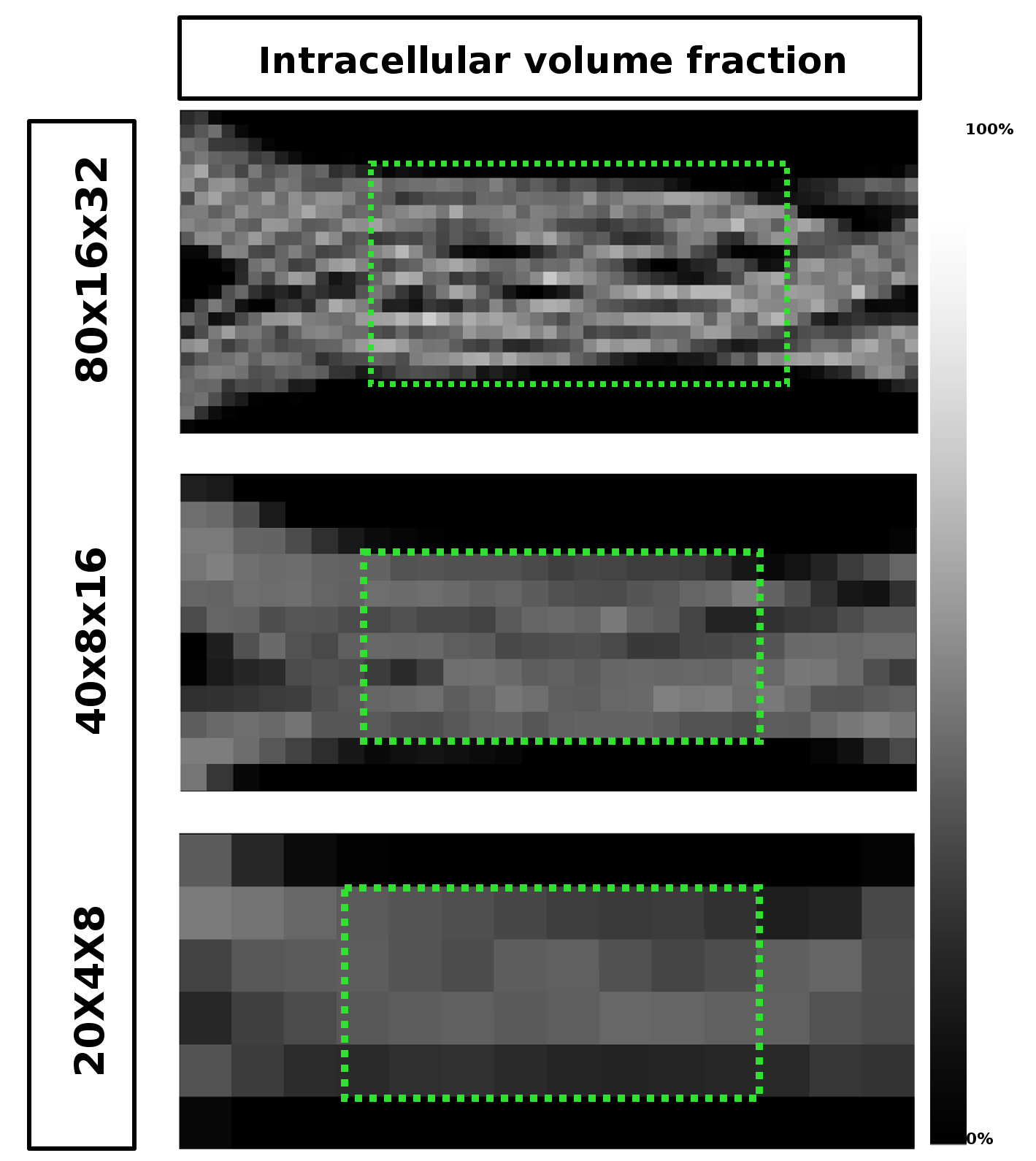}
\caption{The ICVF maps of one volume slice in the XZ-plane in three different resolutions. The highest achieved ICVF value for each resolution were: $0.8013$, $0.5792$, $0.4825$, from top to bottom, respectively. The two green areas highlighted in the two lowest resolutions were used to evaluate the axon diameter estimation.}
\label{fig:ICVF_maps}
\end{figure}

\begin{figure}[h]
\centering
 \textbf{Diameter estimation maps}\par\medskip
\minipage{0.40\textwidth}
\centering
 \textbf{Mean Diameter per voxel}\par\medskip
  \includegraphics[width=\linewidth]{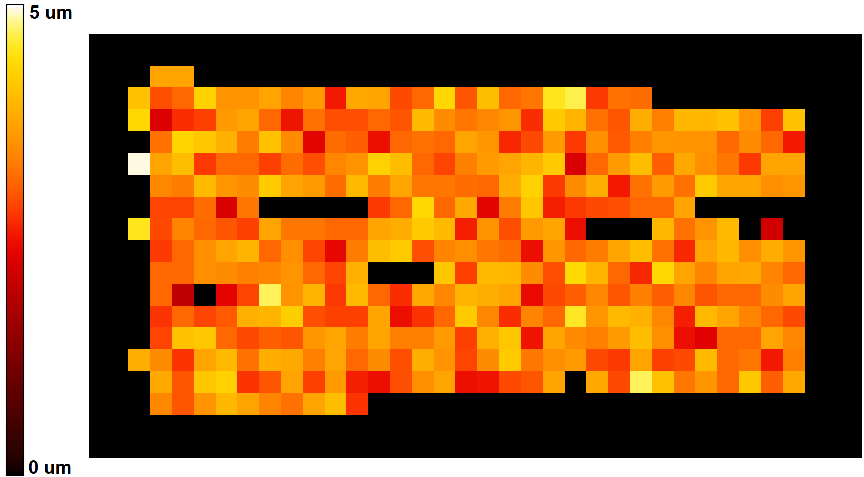}    
\endminipage
\minipage{0.35\textwidth}
\centering
 \textbf{Join estimated diameter histogram  }\par\medskip
  \includegraphics[width=\linewidth]{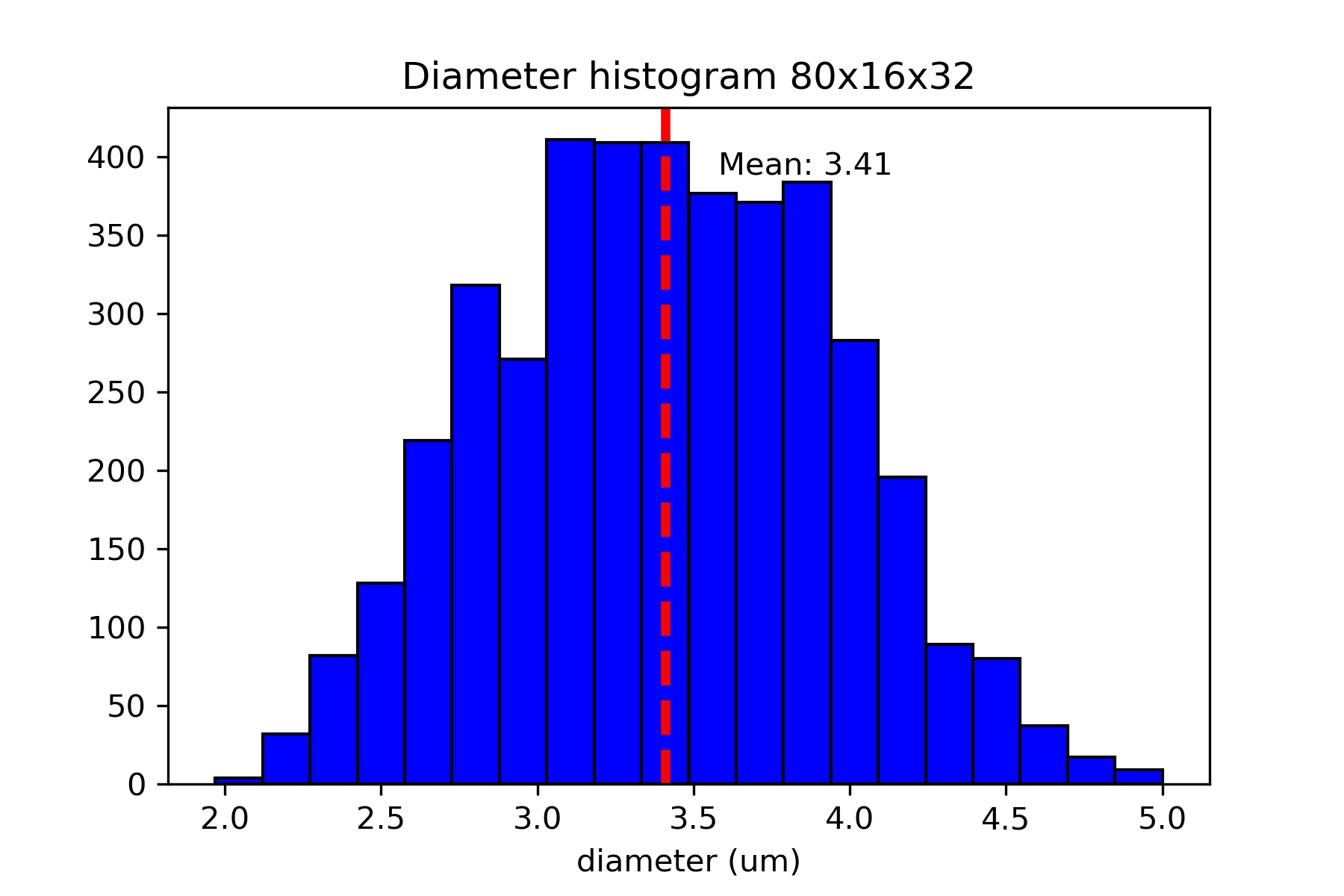}
\endminipage

\minipage{0.40\textwidth}%
\centering
  \includegraphics[width=\linewidth]{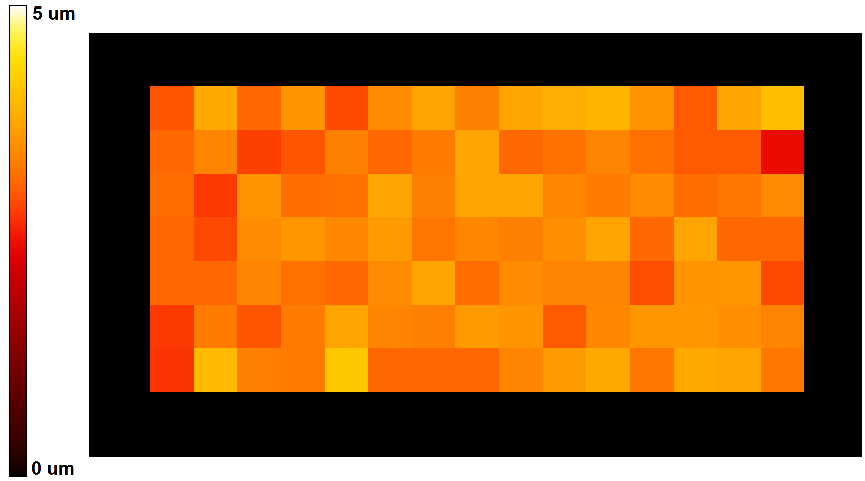}
\endminipage
\minipage{0.35\textwidth}
\centering
  \includegraphics[width=\linewidth]{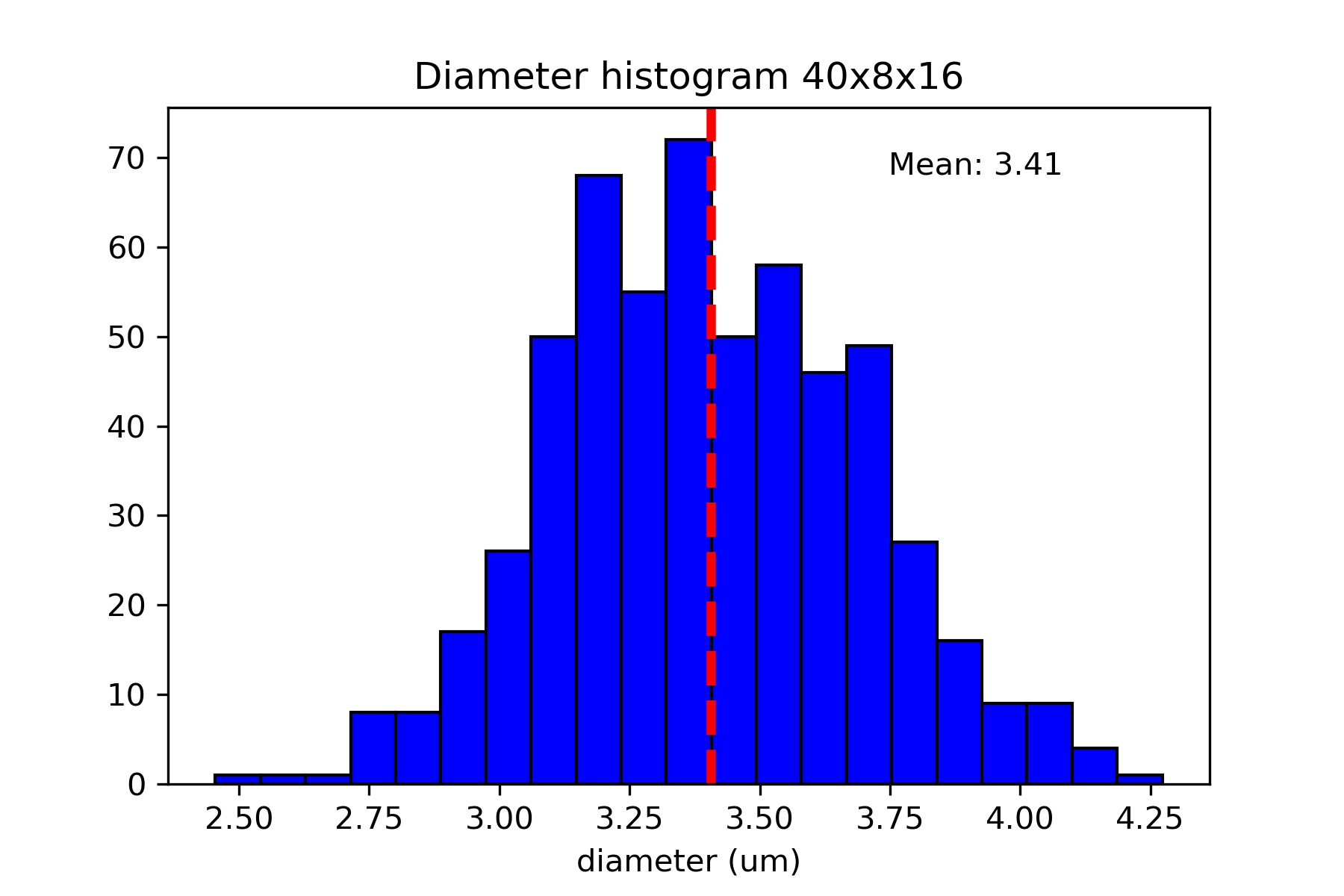}    
\endminipage\hfill

\minipage{0.40\textwidth}
\centering
  \includegraphics[width=\linewidth]{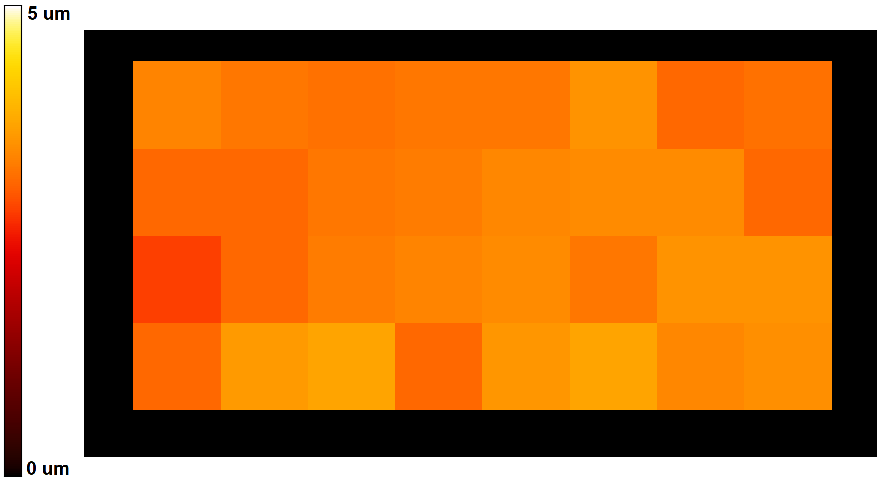}
\endminipage
\minipage{0.35\textwidth}%
\centering
  \includegraphics[width=\linewidth]{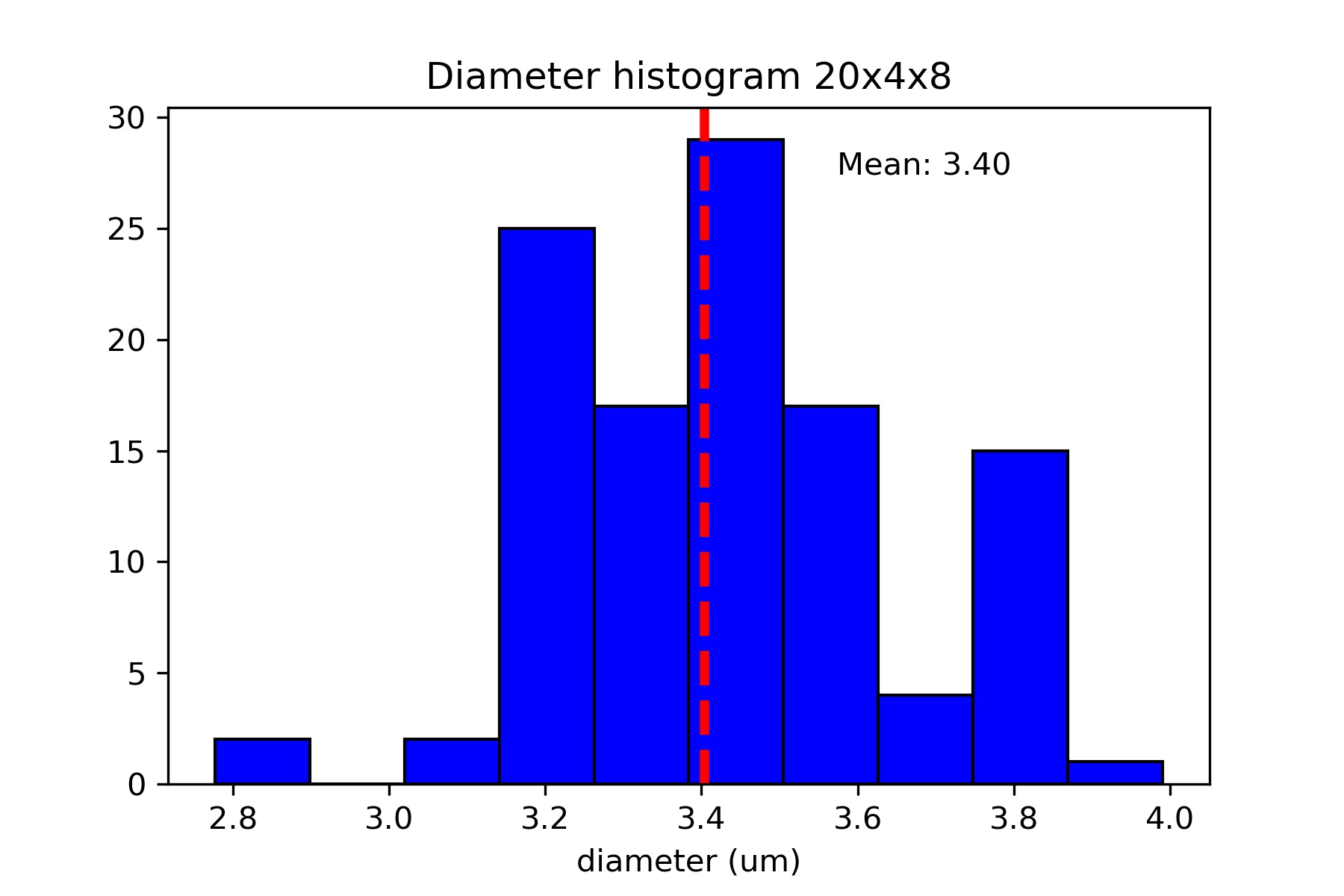}
\endminipage\hfill

\caption{Axon diameter estimation maps (left column) of the regions highlighted in Figure~\ref{fig:ICVF_maps}, and  diameter histograms (right column) estimated on the full volume enclosed by the highlighted regions. Top row shows the axon diameter map and the diameter estimation histogram for the $80 \times 16 \times 32$ nominal resolution; middle row shows the same maps for the $40 \times 8 \times 16$ nominal resolution, and the bottom row shows the same maps for the $20 \times 4 \times 8$ nominal resolution. The dotted line indicates the histograms' mean diameter within the regions, to be compared with the effective apparent diameter ($2*r_{eff}$) of 3.48 um.}
\label{fig:diameter_maps}
\end{figure}

\end{document}